\documentclass[letterpaper,11pt]{book}
\usepackage[top=3cm,bottom=3.22cm,left=4.29cm,right=4.29cm,asymmetric]{geometry}
\usepackage[T1]{fontenc}
\usepackage[utf8]{inputenc}
\usepackage{lmodern}
\usepackage{courier}
\usepackage{csquotes}
\usepackage{tabto}

\usepackage[table,xcdraw]{xcolor}

\usepackage{amssymb,amsmath,amsthm,enumitem,tikz-cd,adjustbox,proof,float,stmaryrd,wasysym,wrapfig,multicol,thmtools,subcaption,pgfplots}
\usetikzlibrary{shapes,snakes,patterns}
\usepackage[11pt]{moresize}

\theoremstyle{definition}

\newtheorem{theorem}{Theorem}[chapter]

\usepackage{scalerel}
\usepackage{mathtools}

\newenvironment{allintypewriter}{\ttfamily}{\par}



\definecolor{gp2green}{RGB}{69, 191, 156}
\definecolor{gp2blue}{RGB}{153, 187, 255}
\definecolor{gp2red}{RGB}{236, 107, 116}
\definecolor{gp2pink}{RGB}{239, 161, 193}
\definecolor{gp2grey}{RGB}{196, 192, 200}
\definecolor{performanceBlue}{RGB}{0, 136, 255}
\definecolor{performanceYellow}{RGB}{252, 199, 17}

\definecolor{dkgreen}{RGB}{0,0.6,0}
\definecolor{grey}{RGB}{0.5,0.5,0.5}
\definecolor{mauve}{RGB}{0.58,0,0.82}

\usepackage{listings}
\usepackage{protobuf/lang}
\usepackage{protobuf/style}
\lstdefinestyle{plain}{
  frame=lines,
  xleftmargin=\parindent,
  belowcaptionskip=1\baselineskip,
  backgroundcolor=\color{proto_background},
  basicstyle=\color{proto_basic}\scriptsize\ttfamily,
  keywordstyle=[1]\color{proto_basic},
  keywordstyle=[2]\color{proto_basic},
  keywordstyle=[3]\color{proto_basic},
  commentstyle=\color{proto_comment},
  stringstyle=\color{proto_string},
  numberstyle=\color{proto_number}\tiny,
  numbers=left,
  numbersep=5pt,
  breaklines=true,
  tabsize=2,
  prebreak=\raisebox{0ex}[0ex][0ex]{\ensuremath{\hookleftarrow}},
  upquote=true,
}

\usetikzlibrary{svg.path}

\definecolor{orcidlogocol}{HTML}{A6CE39}
\tikzset{
  orcidlogo/.pic={
    \fill[orcidlogocol] svg{M256,128c0,70.7-57.3,128-128,128C57.3,256,0,198.7,0,128C0,57.3,57.3,0,128,0C198.7,0,256,57.3,256,128z};
    \fill[white] svg{M86.3,186.2H70.9V79.1h15.4v48.4V186.2z}
                 svg{M108.9,79.1h41.6c39.6,0,57,28.3,57,53.6c0,27.5-21.5,53.6-56.8,53.6h-41.8V79.1z M124.3,172.4h24.5c34.9,0,42.9-26.5,42.9-39.7c0-21.5-13.7-39.7-43.7-39.7h-23.7V172.4z}
                 svg{M88.7,56.8c0,5.5-4.5,10.1-10.1,10.1c-5.6,0-10.1-4.6-10.1-10.1c0-5.6,4.5-10.1,10.1-10.1C84.2,46.7,88.7,51.3,88.7,56.8z};
  }
}

\newcommand\orcidicon[1]{\href{https://orcid.org/#1}{\!\textsuperscript{\mbox{\scalerel*{
\begin{tikzpicture}[yscale=-1,transform shape]
\pic{orcidlogo};
\end{tikzpicture}
}{|}}}}}

\usepackage{hyperref}
\usepackage{graphicx}
\usepackage[british]{babel}
\RequirePackage[style=ieee,backend=bibtex]{biblatex}

\makeatletter
\def\blx@maxline{77}
\makeatother
\makeatletter
\newenvironment{abstract}{
  \null\vfil
  \thispagestyle{empty}
  \@beginparpenalty\@lowpenalty
  \begin{center}%
    \bfseries Abstract \@endparpenalty\@M
  \end{center}%
  }{
  \par\vfil\null
}
\makeatother

\title{\Huge \textbf{Improving the GP\,2 Compiler} \\[0.5cm]}
\author{\large{\textsc{Graham Campbell}\thanks{Supported by a Vacation Internship from the Engineering and Physical Sciences Research Council (EPSRC) in the UK.} \orcidicon{0000-0002-6767-2747}, \textsc{Jack Romo} \orcidicon{0000-0002-8607-7978}, and \textsc{Detlef Plump} \orcidicon{0000-0002-1148-822X}} \\[0.1cm]
\large{\textsc{Department of Computer Science}} \\[0.1cm]
\large{\textsc{University of York, United Kingdom}} \\[0.1cm]
}
\date{September 2019\\~\\Revised February 2020}

\begin{document}

\frontmatter
\maketitle

\clearpage
\thispagestyle{empty}
\cleardoublepage

\begin{abstract}
GP\,2 is an experimental programming language based on graph transformation rules which aims to facilitate program analysis and verification. Writing efficient programs in such a language is hard because graph matching is expensive, however GP\,2 addresses this problem by providing rooted rules which, under mild conditions, can be matched in constant time using the GP\,2 to C compiler. In this report, we document various improvements made to the compiler; most notably the introduction of node lists to improve iteration performance for destructive programs, meaning that binary DAG recognition by reduction need only take linear time where the previous implementation required quadratic time.
\end{abstract}

\clearpage
\thispagestyle{empty}
\cleardoublepage
\phantomsection

\tableofcontents

\cleardoublepage
\phantomsection
\addcontentsline{toc}{chapter}{\listfigurename}
\listoffigures

\chapter{Executive Summary}

GP\,2 is an experimental rule-based programming language based on graph transformation rules which aims to facilitate program analysis and verification. Writing efficient programs in a rule-based language is hard because graph matching is expensive, however GP\,2 addresses this problem by providing rooted rules which, under mild conditions, can be matched in constant time by the code generated by the GP\,2 to C compiler, to which we have:

\begin{enumerate}[itemsep=-0.4ex,topsep=-0.4ex]
\item Introduced node and edge lists to improve iteration performance for destructive programs, allowing the compiler to produce node matching code that skips over an area of deleted nodes in constant time. This allows us to recognise binary DAGs in linear time (given an input graph of bounded degree) by means of a slick reduction algorithm, where in the previous compiler implementation, such a program would have a quadratic worst case time complexity.
\item Fixed parsing input graphs of arbitrary size using Judy arrays. In the previous implementation, one had to specify the maximum number of nodes and edges expected in input host graphs, and the generated code would allocate memory for such a graph on startup, before parsing the input graph. In our updated implementation, it is no longer required to know the maximum input graph size at compile time; we dynamically grow the memory needed during parsing.
\item Added a way to shutdown the generated program as soon as it has written the output graph without cleaning up all the allocated heap memory in a proper way. In general, proper management of the heap memory is a good idea to avoid bugs due to memory leaks or otherwise, however, one might not want to do this in the interest of runtime performance, to allow the process to exit as soon as possible after writing the output.
\item Added a way to ensure matches both reflect root nodes, as well as preserving them. By default, GP\,2 will allow a non-root in a rule LHS to match against either a non-root or a root in a host graph, but if we insist on matches reflecting root nodes, matches can only match non-roots against non-roots. This is useful in order to achieve reversibility of derivations and also to have more control over rule application.
\end{enumerate}

\newpage

We conjecture that the new compiler implementation has worst case time complexity no worse than the original implementation. Moreover, we provide runtime performances of various programs and graph classes as empirical evidence for our conjecture, where timing results differ only by a constant factor. Reduction programs on disconnected graphs are massively improved, changing complexity class, programs that use edge searching heavily, see a runtime performance improvement, such as DFS and transitive closure, and most other programs see a small performance degradation, which we consider worth it for the benefits.

\chapter{Preface}

This report summarises the improvements made to the GP\,2 compiler during August and September 2019. This work was, in part, funded by a Vacation Internship of the Engineering and Physical Sciences Research Council (EPSRC) granted to Graham Campbell. The structure of the report is as follows:

\begin{enumerate}[itemsep=-0.4ex,topsep=-0.4ex]
\item We start by very briefly introduce algebraic graph transformation, the GP\,2 language, and the GP\,2 to C compiler.
\item We will then detail our improvements to the compiler, motivating the changes for each change.
\item Next, we give timing results comparing the original implementation to the improvement implementation, commenting on why it is faster and slower on various classes of program.
\item Finally, we will summarise and evaluate our results.
\end{enumerate}

In the appendices, we have provided usage documentation for current the GP\,2 Compiler, details of the test suite used to give confidence in the correctness of the new implementation, and details of the benchmarking software including the programs used to generate the input graphs.

\mainmatter

\chapter{Introduction} \label{chapter:intro}

In this chapter, we very briefly review the different approaches to algebraic graph transformation, in order to introduce the language GP\,2 and its compiler.


\section{Graph Transformation}

Graph transformation is the rule-based modification of graphs, and is a discipline dating back to the 1970s. There are various approaches to graph transformation, most notably the \enquote{node replacement} \cite{Engelfriet-Rozenberg97a}, \enquote{edge replacement} \cite{Drewes-Kreowski-Habel97a}, and \enquote{algebraic} approaches \cite{Corradini-Montanari-Rossi-Ehrig-Heckel-Lowe97a,Ehrig-Heckel-Korff-Lowe-Ribeiro-Wagner-Corradini97a}, originally developed at the Technical University of Berlin by Ehrig, Pfender, and Schneider \cite{Ehrig-Pfender-Schneider73a,Ehrig79a}. The two major approaches to algebraic graph transformation are the so called \enquote{double pushout} (DPO) approach, and the \enquote{single pushout} (SPO) approach.

Because the DPO approach operates in a structure-preserving manner (rule application in SPO is without an interface graph, so there are no dangling condition checks), this approach is more widely used than the SPO \cite{Ehrig-Heckel-Korff-Lowe-Ribeiro-Wagner-Corradini97a,Ehrig-Ehrig-Prange-Taentzer06a}. Moreover, the DPO approach is genuinely local in the sense that each rule application can only modify the local area of the graph in which it is matched, as opposed to SPO, which allows arbitrary implicit edge deletion. More recently, there have been hybrid approaches that attempt to gain the locality of DPO, but the flexibility of SPO, such as the Sesqui-Pushout approach \cite{Corradini-Heindel-Hermann-Konig06a,Danos-Heindel-HonoratoZimmer-Stucki14a}, which is compatible with DPO when we have injective matches and linear rules \cite{Habel-Muller-Plump01a}.

There are a number of languages and tools, such as
AGG \cite{Runge-Ermel-Taentzer11a},
GMTE \cite{Hannachi-Rodriguez-Drira-Pomares-Saul13a},
Dactl \cite{Glauert-Kennaway-Sleep91a},
GP\,2 \cite{Plump12a},
GReAT \cite{Agrawal-Karsai-Neema-Shi-Vizhanyo06a},
GROOVE \cite{Ghamarian-Mol-Rensink-Zambon-Zimakova12a},
GrGen.Net \cite{Jakumeit-Buchwald-Kroll10a},
Henshin \cite{Arendt-Biermann-Jurack-Krause-Taentzer10a},
PROGRES \cite{Schurr-Winter-Zundorf99a},
and PORGY \cite{Fernandez-Kirchner-Mackie-Pinaud14a}.
It is reasonable that a general purpose local graph transformation language should choose the DPO approach with injective matches and linear rules; GP\,2 is such a language. Moreover, Habel and Plump show that such languages can be \enquote{computationally complete} \cite{Habel-Plump01a}.


\section{The GP\,2 Language}

GP\,2 is an experimental non-deterministic rule-based language for problem solving in the domain of graphs, developed at York, the successor of GP \cite{Plump09a,Plump12a}. GP\,2 is of interest because it has been designed to support formal reasoning on programs \cite{Plump16a}, with a semantics defined in terms of partially labelled graphs, using the injective DPO approach with linear rules and relabelling \cite{Habel-Muller-Plump01a,Habel-Plump02a}.

GP\,2 programs transform input graphs into output graphs, where graphs are directed and may contain parallel edges and loops. Both nodes and edges are labelled with lists consisting of integers and character strings. This includes the special case of items labelled with the empty list. The principal programming construct in GP\,2 consist of conditional graph transformation rules labelled with expressions. Rules operate on \enquote{host graphs} which are labelled with constant values. Formally, the application of a rule to a host graph is defined as a two-stage process in which first the rule is instantiated by replacing all variables with values of the same type, and evaluating all expressions. This yields a standard rule (without expressions) in the DPO approach with relabelling. In the second stage, the instantiated rule is applied to the host graph by constructing two suitable pushouts. The formal semantics of GP\,2 is given in the style of Plotkin's structural operational semantics \cite{Plotkin04a}. Inference rules, first given in \cite{Plump12a}, inductively define a small-step transition relation on configurations. Up-to-date versions can be found in Bak's Thesis \cite{Bak15a}.

Intuitively, applying a rule $L \Rightarrow R$ to a host graph $G$ works as follows: (1) Replace the variables in $L$ and $R$ with constant values and evaluate the expressions in $L$ and $R$, to obtain an instantiated rule $\hat{L} \Rightarrow \hat{R}$. (2) Choose a subgraph $S$ of $G$ isomorphic to $\hat{L}$ such that the dangling condition and the rule's application condition are satisfied (see below). (3) Replace $S$ with $\hat{R}$ as follows: numbered nodes stay in place (possibly relabelled), edges and unnumbered nodes of $\hat{L}$ are deleted, and edges and unnumbered nodes of $\hat{R}$ are inserted.  In this construction, the \enquote{dangling condition} requires that nodes in $S$ corresponding to unnumbered nodes in $\hat{L}$ (which should be deleted) must not be incident with edges outside $S$. The rule's application condition is evaluated after variables have been replaced with the corresponding values of $\hat{L}$, and node identifiers of $L$ with the corresponding identifiers of $S$.

A program consists of declarations of conditional rules and procedures, and exactly one declaration of a main command sequence, which is a distinct procedure named \texttt{Main}. Procedures must be non-recursive, they can be seen as macros. We describe GP\,2's main control constructs. The call of a rule set $\{r_1,\dots,r_n\}$ non-deterministically applies one of the rules whose left-hand graph matches a subgraph of the host graph such that the dangling condition and the rule's application condition are satisfied. The call \enquote{fails} if none of the rules is applicable to the host graph.  The command \texttt{if} $C$ \texttt{then} $P$ \texttt{else} $Q$ is executed on a host graph $G$ by first executing $C$ on a copy of $G$. If this results in a graph, $P$ is executed on the original graph $G$; otherwise, if $C$ fails, $Q$ is executed on $G$. The \texttt{try} command has a similar effect, except that $P$ is executed on the result of $C$'s execution. The loop command $P!$ executes the body $P$ repeatedly until it fails. When this is the case, $P!$ terminates with the graph on which the body was entered for the last time. The \texttt{break} command inside a loop terminates that loop and transfers control to the command following the loop.

Poskitt and Plump have set up the foundations for verification of GP\,2 programs \cite{Poskitt-Plump12a,Poskitt-Plump13a,Poskitt13a,Poskitt-Plump14a} using a Hoare-Style \cite{Hoare69a} system (actually for GP \cite{Manning-Plump08a,Plump09a}), Hristakiev and Plump have developed static analysis for confluence checking \cite{Hristakiev-Plump18a,Hristakiev18a}, and Bak and Plump have extended the language, adding root nodes \cite{Bak-Plump12a,Bak15a}. Plump has shown computational completeness \cite{Plump17a}. Most recently, Atkinson, Plump, and Stepney have developed a probabilistic extension to GP\,2 \cite{Atkinson-Plump-Stepney18a,Atkinson-Plump-Stepney18b}. Most recently, Campbell, Courtehoute and Plump have been interested in linear time algorithms in GP\,2 \cite{Campbell-Courtehoute-Plump19b}, motivating some of the work in this report. We also build on our earlier work in \cite{Campbell-Romo-Plump18a}.


\section{Rooted GP\,2 Programs}

The bottleneck for efficiently implementing algorithms in a language based on graph transformation rules is the cost of graph matching. In general, to match the left-hand graph $L$ of a rule within a host graph $G$ requires time polynomial in the size of $L$ \cite{Bak-Plump12a}. As a consequence, linear-time graph algorithms in imperative languages may be slowed down to polynomial time when they are recast as rule-based programs. To speed up matching, GP\,2 supports \enquote{rooted} graph transformation where graphs in rules and host graphs are equipped with so-called root nodes, originally developed by D\"orr \cite{Dorr95a}. Roots in rules must match roots in the host graph so that matches are restricted to the neighbourhood of the host graph's roots. We draw root nodes using double circles.

A conditional rule \((L \Rightarrow R, c)\) is \enquote{fast} if (1) each node in $L$ is undirectedly reachable from some root, (2) neither $L$ nor $R$ contain repeated occurrences of list, string or atom variables, and (3) the condition $c$ contains neither an $\mathtt{edge}$ predicate nor a test $e_1 = e_2$ or $e_1\, !\!= e_2$ where both $e_1$ and $e_2$ contain a list, string or atom variable. Conditions (2) and (3) will be satisfied by all rules occurring in the following sections; in particular, we neither use the $\mathtt{edge}$ predicate nor the equality tests.

\begin{theorem}[Complexity of matching fast rules \protect\cite{Bak-Plump12a}] \label{thm:rooted-matching-complexity}
Rooted graph matching can be implemented to run in constant time for fast rules, provided there are upper bounds on the maximal node degree and the number of roots in host graphs.
\end{theorem}


\section{Bak's GP\,2 Compiler} \label{sec:bakcompiler}

Before we discuss our modifications to the GP\,2 Compiler\footnote{\url{https://github.com/UoYCS-plasma/GP2}}, we first outline its prior state. The compiler detailed in Bak's Thesis \cite{Bak15a} compiled GP\,2 programs into C code with a Makefile, which was then compiled by the GCC compiler into an executable.

The original compiler stored a graph as a dynamic array of nodes and another of edges, along with the memorised amount of each element. Internally, these arrays were actually more subtle, consisting of an array of the actual elements and a secondary array of indices that contained nothing, or \enquote{holes}. Focusing on nodes, we term the first array the node array and the second the hole array.

In order to iterate through nodes, a program would simply iterate through all indices between 0 and the largest index holding a node, a value the graph remembers. Each index would have to be checked to ensure it genuinely did hold a node, and was not in fact empty. A pointer could be resolved from the index and the node could be modified as desired.

To delete a node, a program would simply add the given index to the array of holes and set the node's entry in the node array to all zeros. Should the node happen to be the last entry in the array, the number of elements in the array could instead be decreased rather than add another hole. Future iterations through the node array would then skip over this hole in the array, as they checked every entry for being a hole or not. This raised performance issues if a program deleted a large number of nodes, for instance, in a graph reduction algorithm, as the enormous number of holes would make traversing the final smaller graph as slow as the original larger one. A prime example of this is the most obvious program that recognises discrete graphs by reduction:

\vspace{0.2em}
\begin{figure}[H]
\centering
\noindent
\fbox{\begin{minipage}{12.6cm}
\begin{allintypewriter}
Main = del!; if node then fail

\medskip
\setlength{\tabcolsep}{16pt}
\vspace{2.5mm}
\begin{tabular}{  p{4.4cm}  p{4.4cm}  }

    \vspace{-1mm} del(x:list) & \vspace{-2mm} node(x:list) \\
    
    \vspace{-2mm}
    \adjustbox{valign=t}{\begin{tikzpicture}[every node/.style={inner sep=0pt, text width=6.5mm, align=center}]
    \node (a) at (0.0,0) [draw,circle,thick] {x};
    \node (b) at (1.0,0) {$\Rightarrow$};
    \node (c) at (2.0,0) [] {$\emptyset$};
    \end{tikzpicture}}
    
    & 

    \vspace{-2mm}
    \adjustbox{valign=t}{\begin{tikzpicture}[every node/.style={inner sep=0pt, text width=6.5mm, align=center}]
    \node (a) at (0.0,0) [draw,circle,thick] {x};
    \node (b) at (1.0,0) {$\Rightarrow$};
    \node (c) at (2.0,0) [draw,circle,thick] {x};

    \node (A) at (0.0,-.52) {\tiny{1}};
    \node (D) at (2.0,-.52) {\tiny{1}};
    \end{tikzpicture}}
    \\
\end{tabular}
\end{allintypewriter}
\end{minipage}
}
\caption{GP\,2 Program \texttt{is-discrete.gp2}}
\label{fig:is-discrete-gp2}
\end{figure}

This program should be expected to run in linear time on unmarked, unrooted graphs, as finding an arbitrary node with no further constraints should be a constant-time operation. Alas, each deleted node adds a new hole at the start of the node array, making the program actually take quadratic time due to having to traverse the holes at each rule match.

\vspace{0.4em}
\begin{figure}[H]
\begin{subfigure}{.5\textwidth}
    \centering
    \begin{tikzpicture}[scale=0.7]
\begin{axis}[
xlabel={Number of nodes in input},
ylabel={Execution time (ms)},
xmin=0, xmax=1100000,
ymin=0, ymax=1000,
legend pos=north west,
ymajorgrids=true,
grid style=dashed,
yticklabel style={/pgf/number format/fixed},
]
\addplot[color=performanceBlue, mark=square*] 
coordinates {
    (100000,114.80)
    (200000,199.31)
    (300000,280.68)
    (400000,367.15)
    (500000,457.26)
    (600000,545.06)
    (700000,636.08)
    (800000,730.74)
    (900000,826.28)
    (1000000,921.32)
};
\addlegendentry{New Impl.}
\end{axis}
\end{tikzpicture}
    \caption{New Implementation}
\end{subfigure}
\begin{subfigure}{.5\textwidth}
    \centering
    \begin{tikzpicture}[scale=0.7]
\begin{axis}[
xlabel={Number of nodes in input},
ylabel={Execution time (ms)},
xmin=0, xmax=110000,
ymin=0, ymax=16000,
legend pos=north west,
ymajorgrids=true,
grid style=dashed,
yticklabel style={/pgf/number format/fixed},
]
\addplot[color=performanceYellow, mark=square*] 
coordinates {
    (10000,128.14)
    (20000,460.44)
    (30000,1017.36)
    (40000,1806.24)
    (50000,2981.68)
    (60000,4460.76)
    (70000,6321.76)
    (80000,8577.40)
    (90000,11436.25)
    (100000,14617.81)
};
\addplot[color=performanceBlue, mark=square*] 
coordinates {
    (10000,20.17)
    (20000,28.81)
    (30000,38.40)
    (40000,44.79)
    (50000,54.45)
    (60000,62.15)
    (70000,70.16)
    (80000,77.73)
    (90000,86.29)
    (100000,94.46)
};
\addlegendentry{Old Impl.}
\addlegendentry{New Impl.}
\end{axis}
\end{tikzpicture}
    \caption{Both Implementations}
\end{subfigure}
\vspace{0.8em}
\caption{Measured Performance of \texttt{is-discrete.gp2}}
\label{fig:is-discrete-timing}
\end{figure}

When inserting a new node, one could simply pop off the last element of the holes array to give a free index to work within. If the holes array was empty, the largest used index would have to be incremented and the node placed at the end of the node array. Should the node array be too small, it would be doubled with the \lstinline{realloc()} C standard library function. The same would be true for the hole array, albeit holes in the hole array would not need to be tracked as only the last element is ever accessed. Arrays would never halve in size should they shrink, to avoid needles memory operations. However, simply reallocating the array raised the possibility of the array changing position in memory to double in size, making any pointers to nodes held while adding a new node invalid. This meant that Bak was indeed required to store indices to nodes instead of pointers, adding extra memory operations to resolve indices every time a node was accessed.

To accommodate root nodes, Bak added a linked list of root nodes to each graph, each entry holding a pointer to a node in the graph's node array. Nodes would contain a flag themselves detailing if they were a root node or not. Root node iteration could then simply be done by traversing this list, a constant time operation should the number of root nodes be constant bounded. Deleting a root node would now require traversing the list as well, to erase the desired entry from said list. These costs are of course a constant time overhead, should the number of root nodes indeed be upper-bounded by a constant in the compiled program.

A node itself would contain a number of entries, including its own index, the number of edges for which it is a target, termed its indegree, the number of edges for which it is a source, termed its outdegree, its incident edges, label, mark and several boolean flags detailing its nature. These flags were stored in separate bool entries in the structure, wasting several bytes where only a bit is used.

The incident edges to a node were stored in a unique manner, with indices of the first two edges being stored statically as part of the node type itself, and indices of all future incoming and outgoing edges being stored in two dynamically allocated arrays of incoming and outgoing edges, respectively. This would avoid having to allocate arrays for a node should its degree be less than three.

Each edge would contain its own index, its label, mark, and the indices of its source and target nodes. It would, similarly to nodes, hold a boolean flag of whether it had been matched or not yet to prevent overlapping matches.

In order to parse an incoming graph, Bak implemented a graph parser in Bison which would accumulate nodes and edges into a graph as it found them in the input. In order to resolve source and target indices, nodes would be stored in a pre-allocated secondary array, stored at the array index equal to the ID the node was given in the input file. When an edge was then discovered, resolving its source and target could be done by looking up the indices in this array equal to the source and target IDs in the input. Unfortunately, this opened a large number of issues: for instance, the array was unable to change its size, meaning only a fixed number of nodes could be entered into the graph before attempting to store nodes at unallocated indices, resulting in unhandled segmentation faults. Also, should a small number of nodes be given enormous IDs, the parsing stage would have to use a huge amount of memory to create an array able to assign at these indices, making this method both incorrect and inefficient should it have been fixed directly.

Finally, to accommodate programs running within the condition of an if statement, for instance, a stack of states was needed. Bak implemented graph stacks in two varieties: copying the previous graph into a stack to reuse it when unwinding, and storing changes to the graph in the stack to undo them in reverse when unwinding. The former simply stored all the data that a node or edge contained, reconstructing the node or edge as needed as it unwound. Graph copying would simply perform a deep copy of the graph as expected.


\section{GP\,2 Compiler Changes}

Since Bak completed his PhD, the GP\,2 Compiler implementation has not stood still. We consider the state of the codebase at the end of 1st September 2015 to be the state of the GP\,2 compiler, as described by Bak's thesis. This is represented by the \texttt{sep-2015} tag on GitHub. Between September 2015 and July 2017 there were various minor fixes and changes made by Bak and Hristakiev.

During the early days of GP\,2, a graphical editor\footnote{\url{https://github.com/UoYCS-plasma/GP2-editor}} was created by Elliot in 2013 \cite{Elliot13a} in C++. Hristakiev worked to revive this implementation in 2015, however the project was abandoned in 2017. Hand has started work on a graph visualiser and visual rule editor\footnote{\url{https://github.com/sdhand/grape}} to run in the web browser, independently of the C compiler implementation \cite{Hand19a}. It is likely that there will eventually be a browser-based GP\,2 editor in the future, possibly based on Hand's work.

Most notably, the following changes to the language concrete syntax were completed before the end of December 2015:

\begin{enumerate}[itemsep=-0.7ex,topsep=-0.7ex]
\item Replaced the syntax for positioning nodes and edges in rules. This change affects the GP\,2 Editor only.
\item String literals can now contain any printable ASCII character.
\item Node and edge identifiers in rules can now start with a digit.
\item Host graph node and edge identifiers now have to be integers.
\end{enumerate}

During the same period, automatic generation of a \texttt{Makefile} by the GP\,2 compiler was added, and so also the option to validate a single rule as input, just like validating an entire program. It's worth noting that the change to allow string literals to contain any printable ASCII character was actually mistakenly only applied to the program concrete syntax definitions. It was not until March 2017 that the grammar for host graphs was updated too.

Bak's last tweak to the GP\,2 Compiler implementation was made in July 2017. Since then, Atkinson, Campbell and Romo have continued to make bug fixes and corrections, without modifying the intended syntax and semantics. Most notably, in July 2018 Romo added the ability to specify as a compiler argument the number of nodes and edges the compiler's generated parser sets aside memory for when parsing host graphs, though this change did not actually become part of the official implementation until February 2019. Previous to that, one had to modify the source code of the compiler itself in order to accept larger input graphs.

During August and September 2019, we have updated the compiler implementation to address the issues described in Section \ref{sec:bakcompiler}. We describe the changes in detail in Chapter \ref{chapter:improvedimpl}. At time of writing, there exist two branches on GitHub: \texttt{legacy} and \texttt{master}. The \texttt{legacy} branch contains the original GP\,2 compiler implementation with support for root reflecting morphisms backported from the new version, which is on the \texttt{master} branch.

Finally, it is worth noting that the implementation of the Probabilistic GP\,2 (P-GP\,2)\footnote{\url{https://github.com/UoYCS-plasma/P-GP2}} \cite{Atkinson-Plump-Stepney18b} Compiler remains separate from the GP\,2 Compiler, was entirely developed by Atkinson, based on the implementation of GP\,2 as it stood after Bak finished making modifications in 2017.

\chapter{Improved Implementation} \label{chapter:improvedimpl}

We now present our modifications to the GP\,2 compiler, to overcome the previous issues highlighted. Our first improvements consisted of fixing graph parsing, abandoning dynamic arrays of nodes in favour of Judy arrays. Our next and most significant internal change is employing linked lists for node and edge iteration, allowing one to jump over deleted elements. In turn, the interface to the compiler was modified to accommodate a range of new optimisation options, adding flags to toggle internal optimisations for each compiled program. We also added a \enquote{fast shutdown mode}, enabling users to choose for their program to terminate without freeing memory, and an option to reflect root nodes in a program if desired. Finally, an integration test suite was built for more efficient and sound compiler development.


\section{Graph Parsing}

To resolve the issues with parsing, we decided to employ Judy arrays\footnote{\url{http://judy.sourceforge.net}} \cite{Silverstein02a}, instead of a simple dynamic array. Invented by Doug Baskins, Judy arrays are a highly cache-optimised hash table implementation. The size of a Judy array is not statically pre-determined and is adjusted, at runtime, to accommodate the number of keys, which themselves can be integers or strings. Instead of storing nodes in the array directly, we also instead store pointers to nodes in the host graph as Judy arrays can only store references to a single word of data. Reallocating the array when doubling it could move the array around and invalidate previous pointers, an issue we resolve in the next section. This allowed an edge to retrieve pointers to its source and target efficiently due to Judy arrays' fast runtime performance \cite{Silverstein02a,Luan-Du-Wang-Ni-Chen07a}. This also resolved problems with unnecessary node array size, allowing node IDs to be arbitrarily large without causing memory problems, as the array simply saw these IDs as meaningless keys in key-value pairs.


\section{From Arrays to Linked Lists} \label{sec:arraystolists}

In order to resolve issues with traversing an array with holes, we decided instead to employ a linked list of nodes, allowing us to jump over any deleted nodes in a single step. We would have new types for entries in the linked list, containing a pointer to the next element and to the current node or edge in question. This would allow us to, for instance, run our discrete graph deletion program in linear time, as the first node in the list could be accessed in constant time.

A performance issue would soon become apparent: should every list element, node and edge not be stored in arrays anymore? Each would have to have a piece of memory allocated for themselves dynamically. Having several calls to \lstinline{malloc()} for every added node and edge would be highly inefficient, requiring many system calls and memory operations. To resolve this, we decided nodes, edges and linked list entries should all be stored in dynamic arrays, doubling in size when too small.

At this stage, it became apparent that node indices were largely redundant, exposing internal implementation too much and adding unneeded memory operations to resolve a node's address every time it was accessed. Thus, we set about rewriting the runtime to use pointers to nodes and edges rather than indices. This in turn added the problem of pointers being invalidated should the array of nodes or edges be moved when \lstinline{realloc()} is called to enlarge them. To resolve this problem in turn, we replaced all internal arrays with a new type we dubbed \texttt{BigArray}s.

The \texttt{BigArray} type, in essence, is an array of pointers to arrays of entries. Each successive array pointed to is double the size of the previous one. The first entry stores two elements, the second four, and so forth. The \texttt{BigArray} type is generic, opting to remember the size of its entries and treating each entry as a chunk of memory rather than deal with actual types of entry. Accessing a given index is constant time, using the position of the largest set bit in the index to identify which sub-array to access. Only a logarithmic number of memory allocations are performed overall, with the overall array of arrays being reallocated $O(log(log(n)))$ times - a trivial amount. Moreover, this is an additive value, not multiplicative.

\texttt{BigArray}s also contain a static chunk of 160 bytes in themselves, allowing for the first few entries in the array to be stored without having to allocate secondary arrays. When nodes use \texttt{BigArray}s of their incident edges, this would mean avoiding unnecessary memory allocations for nodes of small degree, in turn generalising Bak's solution of storing the first two node indices statically with cleaner control logic.

\texttt{BigArray}s also manage holes like the prior implementation did. However, instead of using a second array of holes, \texttt{BigArray}s instead store a linked list of holes within the hole entries in the array, keeping a pointer to the first hole. When a hole is created in the array, that position in the array is overwritten with the data of a new linked list entry, becoming the head of the list of holes. This avoids having to use extra memory for holes, making \texttt{BigArray}s more memory efficient and making deletion of elements constant time.

Most importantly, \texttt{BigArray}s allow one to allocate more memory to the array without having to possibly move previous entries in memory, simply creating a new array. This means the low number of memory allocations may be maintained without pointers to nodes and edges being invalidated.

Thus, three \texttt{BigArray}s are now stored within a graph, one for nodes, one for edges, and one for entries in the linked list of nodes, termed \texttt{NodeList}s. A \texttt{NodeList} simply contains a pointer to the node it refers to and a pointer to the next entry in the linked list. The same is true of \texttt{EdgeList}s, albeit for edges.

Each node now contains a \texttt{BigArray} of linked list entries for edges and pointers to the linked list of outgoing edges and of incoming edges. No iteration through edges directly is ever needed beyond printing a graph, which can be done by iterating through the outgoing edges of every node, so no total list of edges is maintained.

A new issue now presented itself: should a node or edge be deleted, all pointers now represent garbage, the element having possibly been overwritten by a hole. This shed light on yet another possible optimisation: nodes and edges should remember who references them, and be garbage collected when they are referenced by no one. Nodes now retain flags representing if they are in a graph or referred to in the stack of graph changes, and edges remember if they are in a node's list of incoming/outgoing edges or in the stack also. Should a node or edge ever be deleted, the operation can be deferred should other references still exist. Now, stacks of graph changes can simply store pointers to nodes and edges rather than any data in them, stopping them from being truly deleted but simply ignored by the graph that \enquote{deleted} them, with no linked list entry pointing to them. This garbage collection saves on memory in graph stacks and reduces memory operations.

In light of these modifications, it became apparent that graph copying would be somewhat redundant, as stored pointers would resolve to the original graph and not the copied one. We elected to remove graph copying rather than augment copying because of this, only supporting undoing changes now. Also, all flags for nodes and edges were condensed into a single \texttt{char} rather than separate booleans, to save on memory.


\section{Separate Edge Lists} \label{sec:edgelists}

In Section \ref{sec:arraystolists}, we described now the array structure for nodes and edges has changed. In the previous version of the compiler, there was only a single edge array associated with each node. In the new compiler, each node has, in addition to an internal \texttt{BigArray} storing edges, two separate \texttt{EdgeList}s: one for the outgoing edges, and the other for the incoming edges.

It is now possible to run programs that previously required bounded degree to obtain a certain worst case time complexity, now with only bounded incoming degree, or bounded outgoing degree, since search plans can now only consider edges of the correct orientation.


\section{Fast Shutdown Mode} \label{sec:fastshutdown}

It is good practice, in order to aid runtime of analysis of programs for heap memory bugs, for programs to track and cleanup their allocated heap memory. In particular, when doing this, one must accept a one-time cost when cleaning up the data structures after writing the output graph, during the \enquote{shutdown} step. In order to mitigate this cost, we have introduced a way to turn this off, so that the generated code will simply exit the process as soon as printing the graph. The so called \enquote{fast shutdown} mode can be enabled with a flag passed to the compiler, as documented in Appendix \ref{appendix:usage}.

In addition to fast shutdown mode, we have also introduced a more aggressive optimisation designed to improve the performance of reduction programs, that will turn off garbage collection of nodes and edges all together, as well as disabling refcounting of host label lists, making garbage collection of host label lists impossible. Once again, documentation of how to enable this more aggressive optimisation is provided in Appendix \ref{appendix:usage}.


\section{Root Reflecting Mode} \label{sec:rootrefl}

GP\,2 theoretical foundation is rooted graph transformation with relabelling. GP\,2 rule schemata are used to instantiate genuine \enquote{rules} which are then applied in the standard way. Unfortunately, Plump and Bak's model results in derivations not being invertible, even when ignoring application conditions. This is because the right square of a derivation need not be a natural pushout. This has the unfortunate consequence that derivations are not invertible. Moreover, that non-roots can be matched against roots, so a rule that was intended to introduce a new root node, might actually behave like \texttt{skip}.

\vspace{1em}
\begin{figure}[H]
\centering
\noindent
\begin{tikzpicture}[every node/.style={inner sep=0pt, text width=6.5mm, align=center}]
    \node (a) at (0.0,0.0) [draw, circle, thick] {\,};
    \node (b) at (1.0,0.0) {$\leftarrow$};
    \node (c) at (2.0,0.0) [draw, circle, thick] {\,};
    \node (d) at (3.0,0.0) {$\rightarrow$};
    \node (e) at (4.0,0.0) [draw, circle, thick, double, double distance=0.4mm] {\,};

    \node (f) at (0.0,-1.0) {$\big\downarrow$};
    \node (g) at (1.0,-1.0) {NPO};
    \node (h) at (2.0,-1.0) {$\big\downarrow$};
    \node (i) at (3.0,-1.0) {PO};
    \node (j) at (4.0,-1.0) {$\big\downarrow$};

    \node (k) at (0.0,-2.0) [draw, circle, thick, double, double distance=0.4mm] {\,};
    \node (l) at (1.0,-2.0) {$\leftarrow$};
    \node (m) at (2.0,-2.0) [draw, circle, thick, double, double distance=0.4mm] {\,};
    \node (n) at (3.0,-2.0) {$\rightarrow$};
    \node (o) at (4.0,-2.0) [draw, circle, thick, double, double distance=0.4mm] {\,};
\end{tikzpicture}
\vspace{-0.4em}
\caption{Example Rooted Derivation}
\end{figure}
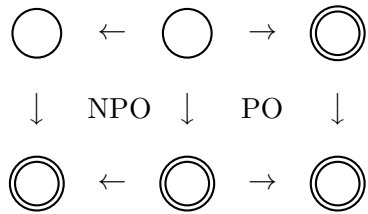

Campbell has proposed a new foundation for rooted GT systems with relabelling \cite{Campbell19a}, that mitigates this problem. Instead of only insisting on matches preserving root nodes, we also insist on them reflecting them too. This was formalised by defining rootedness using a partial function into a two-point set rather than pointing graphs with root nodes, thus allowing both squares in a derivation to be natural pushouts, where rules are allowed to have undefined rootedness in their interface graphs.

In order to simulate this new model of rootedness, one only needs to make small modifications to the compiler to enforce reflection of rootedness nodes in matches. We have thus made this change on both the legacy and master branches of the compiler implementation. For usage details, see Appendix \ref{appendix:usage}.


\section{Link-Time Optimisation}

In order to produce faster compiled programs, we have increased GCC's optimisation flag from \texttt{-O2} to \texttt{-O3}\footnote{\url{https://gcc.gnu.org/onlinedocs/gcc/Optimize-Options.html}}. Moreover, we no longer compile the \enquote{library} files ahead of time for linking. We compile them with the generated program, with \enquote{link-time optimisation}\footnote{\url{https://gcc.gnu.org/wiki/LinkTimeOptimization}} enabled, allowing GCC to more aggressively optimise the whole program.

This has lead to a minor change to the CLI interface of the compiler, and also the generated files. Usage documentation can be found in Appendix \ref{appendix:usage}.


\section{Integration Tests}

In order to have confidence in the correctness of both the \texttt{legacy} and \texttt{master} GP\,2 Compiler, we have written various integration tests. As of 20th September 2019, there are 140 test cases that are checked. Details of the tests can be found in Appendix \ref{appendix:tests}.

\chapter{Timing Results} \label{chapter:timing}

We conjecture that the new compiler implementation has worst case time complexity no worse than the original implementation. We provide runtime performances of various programs and graph classes as empirical evidence for our conjecture, where timing results differ only by a constant factor. We look at the performance of generation programs, reduction programs, and a couple of other programs, including undirected DFS. We show that there are significant improvements for some types of reduction programs.

Details of the benchmarking software and graph generation can be found in Appendix \ref{appendix:benchmarking}, including which compiler flags were used.


\section{Reduction Performance}

\vspace{-0.5em}
\noindent
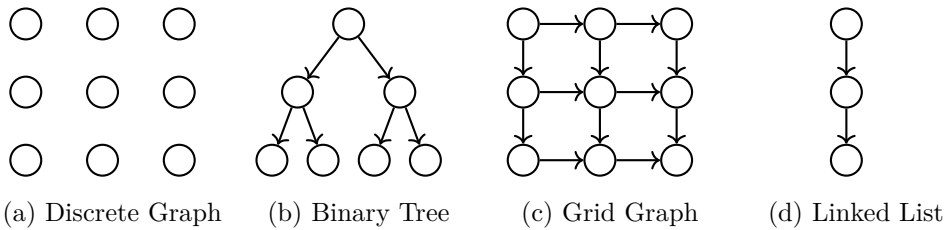
\begin{figure}[H]
\begin{subfigure}{.24\textwidth}
    \centering
    \begin{tikzpicture}[scale=0.68]
    \node (a) at (-1.500,1.333)  [draw,circle,thick] {\,};
    \node (b) at (0.000,1.333)   [draw,circle,thick] {\,};
    \node (c) at (1.500,1.333)   [draw,circle,thick] {\,};
    \node (d) at (-1.500,0.000)  [draw,circle,thick] {\,};
    \node (e) at (0.000,0.000)   [draw,circle,thick] {\,};
    \node (f) at (1.500,0.000)   [draw,circle,thick] {\,};
    \node (g) at (-1.500,-1.333) [draw,circle,thick] {\,};
    \node (h) at (0.000,-1.333)  [draw,circle,thick] {\,};
    \node (i) at (1.500,-1.333)  [draw,circle,thick] {\,};
\end{tikzpicture}
    \vspace{0.2em}
    \caption{Discrete Graph}
\end{subfigure}
\begin{subfigure}{.25\textwidth}
    \centering
    \begin{tikzpicture}[scale=0.68]
    \node (a) at (0.000,1.333)   [draw,circle,thick] {\,};
    \node (b) at (1.000,0.000)   [draw,circle,thick] {\,};
    \node (c) at (-1.000,0.000)  [draw,circle,thick] {\,};
    \node (d) at (1.500,-1.333)  [draw,circle,thick] {\,};
    \node (e) at (0.500,-1.333)  [draw,circle,thick] {\,};
    \node (f) at (-0.500,-1.333) [draw,circle,thick] {\,};
    \node (g) at (-1.500,-1.333) [draw,circle,thick] {\,};
    
    \draw (a) edge[->, thick] (b)
          (a) edge[->, thick] (c)
          (b) edge[->, thick] (d)
          (b) edge[->, thick] (e)
          (c) edge[->, thick] (f)
          (c) edge[->, thick] (g);
\end{tikzpicture}
    \vspace{0.2em}
    \caption{Binary Tree}
\end{subfigure}
\begin{subfigure}{.25\textwidth}
    \centering
    \begin{tikzpicture}[scale=0.68]
    \node (a) at (-1.500,1.333)  [draw,circle,thick] {\,};
    \node (b) at (0.000,1.333)   [draw,circle,thick] {\,};
    \node (c) at (1.500,1.333)   [draw,circle,thick] {\,};
    \node (d) at (-1.500,0.000)  [draw,circle,thick] {\,};
    \node (e) at (0.000,0.000)   [draw,circle,thick] {\,};
    \node (f) at (1.500,0.000)   [draw,circle,thick] {\,};
    \node (g) at (-1.500,-1.333) [draw,circle,thick] {\,};
    \node (h) at (0.000,-1.333)  [draw,circle,thick] {\,};
    \node (i) at (1.500,-1.333)  [draw,circle,thick] {\,};
    
    \draw (a) edge[->, thick] (b)
          (a) edge[->, thick] (d)
          (b) edge[->, thick] (c)
          (b) edge[->, thick] (e)
          (c) edge[->, thick] (f)
          (d) edge[->, thick] (e)
          (d) edge[->, thick] (g)
          (e) edge[->, thick] (f)
          (e) edge[->, thick] (h)
          (f) edge[->, thick] (i)
          (g) edge[->, thick] (h)
          (h) edge[->, thick] (i);
\end{tikzpicture}
    \vspace{0.2em}
    \caption{Grid Graph}
\end{subfigure}
\begin{subfigure}{.24\textwidth}
    \centering
    \begin{tikzpicture}[scale=0.68]
    \node (a) at (0.000,1.333)  [draw,circle,thick] {\,};
    \node (b) at (0.000,0.000)  [draw,circle,thick] {\,};
    \node (c) at (0.000,-1.333) [draw,circle,thick] {\,};
    
    \draw (a) edge[->, thick] (b)
          (b) edge[->, thick] (c);
\end{tikzpicture}
    \vspace{0.2em}
    \caption{Linked List}
\end{subfigure}
\vspace{0.8em}
\caption{Input Graph Classes}
\label{fig:input-graph-types}
\end{figure}

In Section \ref{sec:bakcompiler}, we observed that even the program that simply deleted all isolated nodes could not be executed in linear time by the old implementation. Somewhat more subtle, is when an input graph is connected, however the program splits up the graph as it executes. It is possible to recognise binary DAGs using a slick reduction program (Figure \ref{fig:is-bin-dag-gp2}) with this property. Once again, we observe that the original compiler produces a program that runs in quadratic time on many graph classes, including full binary trees and grid graphs (Figure \ref{fig:input-graph-types}). Our new compiler runs in linear time on such graphs.

\vspace{0.2em}
\begin{figure}[H]
\centering
\noindent
\scalebox{.73}{\fbox{\begin{minipage}{17.5cm}
\begin{allintypewriter}
Main = (init; Reduce!; if flag then break)!; if flag then fail

Reduce = up!; try Delete else set\_flag

Delete = \{del0, del1, del1\_d, del21, del21\_d, del22, del22\_d\}

\medskip
\setlength{\tabcolsep}{16pt}
\vspace{2.5mm}
\begin{tabular}{  p{3.5cm}  p{6.5cm}  p{3.8cm}  }
    
    \vspace{-1mm} init(x:list) & \vspace{-2mm} up(a,x,y:list) & \vspace{-2mm} del0(x:list) \\

    \vspace{-2mm}
    \adjustbox{valign=t}{\begin{tikzpicture}[every node/.style={inner sep=0pt, text width=6.5mm, align=center}]
        \node (a) at (0.0,0) [draw,circle,thick] {x};

        \node (b) at (1.0,0) {$\Rightarrow$};
        
        \node (c) at (2.0,0) [draw,circle,thick,double,double distance=0.4mm] {x};
        
        \node (A) at (0.0,-.52) {\tiny{1}};
        \node (C) at (2.0,-.52) {\tiny{1}};
    \end{tikzpicture}}

    &

    \vspace{-2mm}
    \adjustbox{valign=t}{\begin{tikzpicture}[every node/.style={inner sep=0pt, text width=6.5mm, align=center}]
        \node (a) at (0.0,0) [draw,circle,thick,double,double distance=0.4mm] {x};
        \node (b) at (1.5,0) [draw,circle,thick] {y};
        
        \node (c) at (2.5,0) {$\Rightarrow$};
        
        \node (d) at (3.5,0) [draw,circle,thick] {x};
        \node (e) at (5.0,0) [draw,circle,thick,double,double distance=0.4mm] {y};
        
        \node (A) at (0.0,-.52) {\tiny{1}};
        \node (B) at (1.5,-.52) {\tiny{2}};
        \node (D) at (3.5,-.52) {\tiny{1}};
        \node (E) at (5.0,-.52) {\tiny{2}};
        
        \draw (b) edge[->,thick] node[above, yshift=2.5pt] {a} (a)
              (e) edge[->,thick,dashed] node[above, yshift=2.5pt] {a} (d);
    \end{tikzpicture}}

    &

    \vspace{-2mm}
    \adjustbox{valign=t}{\begin{tikzpicture}[every node/.style={inner sep=0pt, text width=6.5mm, align=center}]
        \node (a) at (0.0,0) [draw,circle,thick,double,double distance=0.4mm] {x};

        \node (b) at (1.0,0) {$\Rightarrow$};
        
        \node (c) at (2.0,0) {$\emptyset$};
    \end{tikzpicture}}
    \\
\end{tabular}
\begin{tabular}{  p{6.9cm}  p{6.9cm}  }

    \vspace{-1mm} del1(a,x,y:list) & \vspace{-2mm} del1\_d(a,x,y:list) \\

    \vspace{-2mm}
    \adjustbox{valign=t}{\begin{tikzpicture}[every node/.style={inner sep=0pt, text width=6.5mm, align=center}]
        \node (a) at (0.0,0)     [draw,circle,thick] {x};
        \node (b) at (1.5,0)     [draw,circle,thick,double,double distance=0.4mm] {y};

        \node (c) at (2.5,0)     {$\Rightarrow$};

        \node (d) at (3.5,0)     [draw,circle,thick,double,double distance=0.4mm] {x};

        \node (A) at (0.0,-.52) {\tiny{1}};
        \node (D) at (3.5,-.52) {\tiny{1}};

        \draw (b) edge[->,thick] node[above, yshift=2.5pt] {a} (a);
    \end{tikzpicture}}

    &

    \vspace{-2mm}
    \adjustbox{valign=t}{\begin{tikzpicture}[every node/.style={inner sep=0pt, text width=6.5mm, align=center}]
        \node (a) at (0.0,0)     [draw,circle,thick] {x};
        \node (b) at (1.5,0)     [draw,circle,thick,double,double distance=0.4mm] {y};

        \node (c) at (2.5,0)     {$\Rightarrow$};

        \node (d) at (3.5,0)     [draw,circle,thick,double,double distance=0.4mm] {x};

        \node (A) at (0.0,-.52) {\tiny{1}};
        \node (D) at (3.5,-.52) {\tiny{1}};

        \draw (b) edge[->,thick,dashed] node[above, yshift=2.5pt] {a} (a);
    \end{tikzpicture}}
    \\
\end{tabular}
\begin{tabular}{  p{6.9cm}  p{6.9cm}  }

    \vspace{-1mm} del21(a,b,x,y:list) & \vspace{-2mm} del21\_d(a,b,x,y:list) \\

    \vspace{-2mm}
    \adjustbox{valign=t}{\begin{tikzpicture}[every node/.style={inner sep=0pt, text width=6.5mm, align=center}]
        \node (a) at (0.0,0)     [draw,circle,thick] {x};
        \node (b) at (1.5,0)     [draw,circle,thick,double,double distance=0.4mm] {y};

        \node (c) at (2.5,0)     {$\Rightarrow$};

        \node (d) at (3.5,0)     [draw,circle,thick,double,double distance=0.4mm] {x};

        \node (A) at (0.0,-.52) {\tiny{1}};
        \node (C) at (3.5,-.52) {\tiny{1}};

        \draw (b) edge[->,thick,bend left=-25] node[above, yshift=2.5pt] {a} (a)
              (b) edge[->,thick,bend left=25] node[above, yshift=2.5pt] {b} (a);
    \end{tikzpicture}}

    &

    \vspace{-2mm}
    \adjustbox{valign=t}{\begin{tikzpicture}[every node/.style={inner sep=0pt, text width=6.5mm, align=center}]
        \node (a) at (0.0,0)     [draw,circle,thick] {x};
        \node (b) at (1.5,0)     [draw,circle,thick,double,double distance=0.4mm] {y};

        \node (c) at (2.5,0)     {$\Rightarrow$};

        \node (d) at (3.5,0)     [draw,circle,thick,double,double distance=0.4mm] {x};

        \node (A) at (0.0,-.52) {\tiny{1}};
        \node (C) at (3.5,-.52) {\tiny{1}};

        \draw (b) edge[->,thick,dashed,bend left=-25] node[above, yshift=2.5pt] {a} (a)
              (b) edge[->,thick,bend left=25] node[above, yshift=2.5pt] {b} (a);
    \end{tikzpicture}}
    \\
\end{tabular}
\begin{tabular}{  p{6.9cm}  p{6.9cm}  }

    \vspace{-1mm} del22(a,b,x,y,z:list) & \vspace{-2mm} del22\_d(a,b,x,y,z:list) \\

    \vspace{-2mm}
    \adjustbox{valign=t}{\begin{tikzpicture}[every node/.style={inner sep=0pt, text width=6.5mm, align=center}]
        \node (a) at (0.0,-0.0)  [draw,circle,thick] {x};
        \node (b) at (0.0,-1.2)  [draw,circle,thick] {y};
        \node (c) at (1.5,-0.6)  [draw,circle,thick,double,double distance=0.4mm] {z};

        \node (d) at (2.5,-0.6)  {$\Rightarrow$};

        \node (e) at (3.5,-0.0)  [draw,circle,thick,double,double distance=0.4mm] {x};
        \node (f) at (3.5,-1.2)  [draw,circle,thick] {y};

        \node (A) at (0.0,-0.52) {\tiny{1}};
        \node (B) at (0.0,-1.72) {\tiny{2}};
        \node (E) at (3.5,-0.52) {\tiny{1}};
        \node (F) at (3.5,-1.72) {\tiny{2}};

        \draw (c) edge[->,thick] node[above, yshift=2.5pt] {a} (a)
              (c) edge[->,thick] node[above, yshift=2.5pt] {b} (b);
    \end{tikzpicture}}

    &

    \vspace{-2mm}
    \adjustbox{valign=t}{\begin{tikzpicture}[every node/.style={inner sep=0pt, text width=6.5mm, align=center}]
        \node (a) at (0.0,-0.00) [draw,circle,thick] {x};
        \node (b) at (0.0,-1.2)  [draw,circle,thick] {y};
        \node (c) at (1.5,-0.6)  [draw,circle,thick,double,double distance=0.4mm] {z};

        \node (d) at (2.5,-0.6)  {$\Rightarrow$};

        \node (e) at (3.5,-0.0)  [draw,circle,thick,double,double distance=0.4mm] {x};
        \node (f) at (3.5,-1.2)  [draw,circle,thick] {y};

        \node (A) at (0.0,-0.52) {\tiny{1}};
        \node (B) at (0.0,-1.72) {\tiny{2}};
        \node (E) at (3.5,-0.52) {\tiny{1}};
        \node (F) at (3.5,-1.72) {\tiny{2}};

        \draw (c) edge[->,thick,dashed] node[above, yshift=2.5pt] {a} (a)
              (c) edge[->,thick] node[above, yshift=2.5pt] {b} (b);
    \end{tikzpicture}}
    \\
\end{tabular}
\begin{tabular}{  p{6.9cm}  p{6.9cm}  }

    \vspace{-1mm} set\_flag(x:list) & \vspace{-2mm} flag(x:list) \\

    \vspace{-2mm}
    \adjustbox{valign=t}{\begin{tikzpicture}[every node/.style={inner sep=0pt, text width=6.5mm, align=center}]
        \node (a) at (0.0,0) [draw,circle,thick,double,double distance=0.4mm] {x};

        \node (b) at (1.0,0) {$\Rightarrow$};

        \node (c) at (2.0,0) [draw,circle,fill=gp2grey,thick,double,double distance=0.4mm] {x};

        \node (A) at (0.0,-.52) {\tiny{1}};
        \node (C) at (2.0,-.52) {\tiny{1}};
    \end{tikzpicture}}

    &

    \vspace{-2mm}
    \adjustbox{valign=t}{\begin{tikzpicture}[every node/.style={inner sep=0pt, text width=6.5mm, align=center}]
        \node (a) at (0.0,0) [draw,circle,fill=gp2grey,thick,double,double distance=0.4mm] {x};

        \node (b) at (1.0,0) {$\Rightarrow$};

        \node (c) at (2.0,0) [draw,circle,fill=gp2grey,thick,double,double distance=0.4mm] {x};

        \node (A) at (0.0,-.52) {\tiny{1}};
        \node (C) at (2.0,-.52) {\tiny{1}};
    \end{tikzpicture}}
    \\
\end{tabular}
\end{allintypewriter}
\end{minipage}
}}
\caption{GP\,2 Program \texttt{is-bin-dag.gp2}}
\label{fig:is-bin-dag-gp2}
\end{figure}

\vspace{0.4em}
\begin{figure}[H]
\begin{subfigure}{.5\textwidth}
    \centering
    \begin{tikzpicture}[scale=0.7]
\begin{axis}[
xlabel={Number of nodes in input},
ylabel={Execution time (ms)},
xmin=0, xmax=140000,
ymin=0, ymax=14000,
legend pos=north west,
ymajorgrids=true,
grid style=dashed,
yticklabel style={/pgf/number format/fixed},
]
\addplot[color=performanceYellow, mark=square*] 
coordinates {
    (8191,62.22)
    (16383,183.16)
    (32767,642.28)
    (65535,2765.78)
    (131071,13657.44)
};
\addplot[color=performanceBlue, mark=square*] 
coordinates {
    (8191,35.65)
    (16383,40.81)
    (32767,69.82)
    (65535,124.63)
    (131071,231.38)
};
\addlegendentry{Old Impl.}
\addlegendentry{New Impl.}
\end{axis}
\end{tikzpicture}
    \caption{Tree Reduction}
\end{subfigure}
\begin{subfigure}{.5\textwidth}
    \centering
    \begin{tikzpicture}[scale=0.7]
\begin{axis}[
xlabel={Number of nodes in input},
ylabel={Execution time (ms)},
xmin=0, xmax=1100000,
ymin=0, ymax=3250,
legend pos=north west,
ymajorgrids=true,
grid style=dashed,
yticklabel style={/pgf/number format/fixed},
]
\addplot[color=performanceYellow, mark=square*] 
coordinates {
    (90000,275.13)
    (160000,444.56)
    (250000,685.72)
    (360000,983.92)
    (490000,1316.96)
    (640000,1742.48)
    (810000,2199.17)
    (1000000,2709.24)
};
\addplot[color=performanceBlue, mark=square*] 
coordinates {
    (90000,293.67)
    (160000,483.92)
    (250000,726.04)
    (360000,1063.11)
    (490000,1448.70)
    (640000,1897.87)
    (810000,2405.17)
    (1000000,2957.17)
};
\addlegendentry{Old Impl.}
\addlegendentry{New Impl.}
\end{axis}
\end{tikzpicture}
    \caption{Grid Reduction}
\end{subfigure}
\vspace{0.8em}
\caption{Measured Performance of \texttt{is-bin-dag.gp2}}
\label{fig:is-bin-dag-timing}
\end{figure}

Next, we look at a rooted tree reduction program by Campbell \cite{Campbell19a} (modified the program to work with unmarked graphs) that was linear time on graphs of bounded degree in the previous implementation of the compiler. We confirm that it remains linear time in the new compiler, and include grid graphs as a negative case for the program to determine they are not trees.

\vspace{0.2em}
\begin{figure}[H]
\centering
\noindent
\scalebox{.73}{\fbox{\begin{minipage}{17.5cm}
\begin{allintypewriter}
Main = init; Reduce!; Unmark; if Check then fail

Reduce = \{prune0, prune1, push\}

Unmark = try unmark; try unmark

Check = \{two\_nodes, has\_loop\}

\medskip
\setlength{\tabcolsep}{16pt}
\vspace{2.5mm}
\begin{tabular}{  p{4cm}  p{4.9cm}  p{4.9cm}  }
    
    \vspace{-1mm} init(x:list) & \vspace{-2mm} prune0(a,x,y:list) & \vspace{-2mm} prune1(a,x,y:list) \\

    \vspace{-2mm}
    \adjustbox{valign=t}{\begin{tikzpicture}[every node/.style={inner sep=0pt, text width=6.5mm, align=center}]
        \node (a) at (0.0,0) [draw,circle,thick] {x};

        \node (b) at (1.0,0) {$\Rightarrow$};
        
        \node (c) at (2.0,0) [draw,circle,thick,double,double distance=0.4mm] {x};
        
        \node (A) at (0.0,-.52) {\tiny{1}};
        \node (C) at (2.0,-.52) {\tiny{1}};
    \end{tikzpicture}}

    &

    \vspace{-2mm}
    \adjustbox{valign=t}{\begin{tikzpicture}[every node/.style={inner sep=0pt, text width=6.5mm, align=center}]
        \node (a) at (0.0,0) [draw,circle,thick] {x};
        \node (b) at (1.5,0) [draw,circle,thick,double,double distance=0.4mm] {y};
        
        \node (c) at (2.5,0) {$\Rightarrow$};
        
        \node (d) at (3.5,0) [draw,circle,thick,double,double distance=0.4mm] {x};
        
        \node (A) at (0.0,-.52) {\tiny{1}};
        \node (D) at (3.5,-.52) {\tiny{1}};
        
        \draw (a) edge[->,thick] node[above, yshift=2.5pt] {a} (b);
    \end{tikzpicture}}

    &

    \vspace{-2mm}
    \adjustbox{valign=t}{\begin{tikzpicture}[every node/.style={inner sep=0pt, text width=6.5mm, align=center}]
        \node (a) at (0.0,0) [draw,circle,fill=gp2grey,thick] {x};
        \node (b) at (1.5,0) [draw,circle,thick,double,double distance=0.4mm] {y};
        
        \node (c) at (2.5,0) {$\Rightarrow$};
        
        \node (d) at (3.5,0) [draw,circle,thick,double,double distance=0.4mm] {x};
        
        \node (A) at (0.0,-.52) {\tiny{1}};
        \node (D) at (3.5,-.52) {\tiny{1}};
        
        \draw (a) edge[->,thick] node[above, yshift=2.5pt] {a} (b);
    \end{tikzpicture}}
    \\
\end{tabular}
\begin{tabular}{  p{4cm}  p{9.8cm}  }

    \vspace{-1mm} unmark(x:list) & \vspace{-2mm} push(a,x,y:list) \\

    \vspace{-2mm}
    \adjustbox{valign=t}{\begin{tikzpicture}[every node/.style={inner sep=0pt, text width=6.5mm, align=center}]
        \node (a) at (0.0,0) [draw,circle,fill=gp2grey,thick] {x};

        \node (b) at (1.0,0) {$\Rightarrow$};
        
        \node (c) at (2.0,0) [draw,circle,thick] {x};
        
        \node (A) at (0.0,-.52) {\tiny{1}};
        \node (C) at (2.0,-.52) {\tiny{1}};
    \end{tikzpicture}}

    &

    \vspace{-2mm}
    \adjustbox{valign=t}{\begin{tikzpicture}[every node/.style={inner sep=0pt, text width=6.5mm, align=center}]
        \node (a) at (0.0,0)     [draw,circle,thick,double,double distance=0.4mm] {x};
        \node (b) at (1.5,0)     [draw,circle,thick] {y};
        
        \node (c) at (2.5,0)     {$\Rightarrow$};
        
        \node (d) at (3.5,0)     [draw,circle,fill=gp2grey,thick] {x};
        \node (e) at (5,0)       [draw,circle,thick,double,double distance=0.4mm] {y};
        
        \node (A) at (0.0,-.52) {\tiny{1}};
        \node (B) at (1.5,-.52) {\tiny{2}};
        \node (D) at (3.5,-.52) {\tiny{1}};
        \node (E) at (5.0,-.52) {\tiny{2}};
        
        \draw (a) edge[->,thick] node[above, yshift=2.5pt] {a} (b)
              (d) edge[->,thick] node[above, yshift=2.5pt] {a} (e);
    \end{tikzpicture}}
    \\
\end{tabular}
\begin{tabular}{  p{4cm}  p{9.8cm}  }

    \vspace{-1mm} has\_loop(a,x:list) & \vspace{-2mm} two\_nodes(x,y:list) \\

    \vspace{-2mm}
    \adjustbox{valign=t}{\begin{tikzpicture}[every node/.style={inner sep=0pt, text width=6.5mm, align=center}]
        \node (a) at (0.0,0) [draw,circle,thick] {x};
        
        \node (b) at (1.0,0) {$\Rightarrow$};
        
        \node (c) at (2.0,0) [draw,circle,thick] {x};
        
        \node (A) at (0.0,-.52) {\tiny{1}};
        \node (C) at (2.0,-.52) {\tiny{1}};
        
        \draw (a) edge[->,in=-30,out=-60,loop,thick] node[right, yshift=1.5pt] {a} (a)
              (c) edge[->,in=-30,out=-60,loop,thick] node[right, yshift=1.5pt] {a} (c);
    \end{tikzpicture}}

    &

    \vspace{-2mm}
    \adjustbox{valign=t}{\begin{tikzpicture}[every node/.style={inner sep=0pt, text width=6.5mm, align=center}]
        \node (a) at (0.0,0) [draw, circle,thick] {x};
        \node (b) at (1.5,0) [draw, circle,thick] {y};

        \node (c) at (2.5,0) {$\Rightarrow$};

        \node (d) at (3.5,0) [draw,circle,thick] {x};
        \node (e) at (5.0,0) [draw,circle,thick] {y};

        \node (A) at (0.0,-.52) {\tiny{1}};
        \node (B) at (1.5,-.52) {\tiny{2}};
        \node (D) at (3.5,-.52) {\tiny{1}};
        \node (E) at (5.0,-.52) {\tiny{2}};
    \end{tikzpicture}}
    \\
\end{tabular}
\end{allintypewriter}
\end{minipage}
}}
\caption{GP\,2 Program \texttt{is-tree.gp2}}
\label{fig:is-tree-gp2}
\end{figure}

\vspace{0.4em}
\begin{figure}[H]
\begin{subfigure}{.5\textwidth}
    \centering
    \begin{tikzpicture}[scale=0.7]
\begin{axis}[
xlabel={Number of nodes in input},
ylabel={Execution time (ms)},
xmin=0, xmax=1100000,
ymin=0, ymax=1800,
legend pos=north west,
ymajorgrids=true,
grid style=dashed,
yticklabel style={/pgf/number format/fixed},
]
\addplot[color=performanceYellow, mark=square*] 
coordinates {
    (65535,122.52)
    (131071,214.20)
    (262143,375.95)
    (524287,713.55)
    (1048575,1404.75)
};
\addplot[color=performanceBlue, mark=square*] 
coordinates {
    (65535,133.40)
    (131071,234.47)
    (262143,440.12)
    (524287,851.67)
    (1048575,1698.62)
};
\addlegendentry{Old Impl.}
\addlegendentry{New Impl.}
\end{axis}
\end{tikzpicture}
    \caption{Tree Reduction}
\end{subfigure}
\begin{subfigure}{.5\textwidth}
    \centering
    \begin{tikzpicture}[scale=0.7]
\begin{axis}[
xlabel={Number of nodes in input},
ylabel={Execution time (ms)},
xmin=0, xmax=1100000,
ymin=0, ymax=2200,
legend pos=north west,
ymajorgrids=true,
grid style=dashed,
yticklabel style={/pgf/number format/fixed},
]
\addplot[color=performanceYellow, mark=square*] 
coordinates {
    (90000,206.53)
    (160000,310.22)
    (250000,466.51)
    (360000,590.19)
    (490000,794.06)
    (640000,1033.59)
    (810000,1303.68)
    (1000000,1599.73)
};
\addplot[color=performanceBlue, mark=square*] 
coordinates {
    (90000,289.97)
    (160000,382.21)
    (250000,563.41)
    (360000,738.82)
    (490000,992.92)
    (640000,1293.69)
    (810000,1636.45)
    (1000000,2018.30)
};
\addlegendentry{Old Impl.}
\addlegendentry{New Impl.}
\end{axis}
\end{tikzpicture}
    \caption{Grid Reduction}
\end{subfigure}
\vspace{0.8em}
\caption{Measured Performance of \texttt{is-tree.gp2}}
\label{fig:is-tree-timing}
\end{figure}

Finally, we look at recognition of Series-Parallel graphs \cite{Duffin65a,Plump09a,Plump16a}. We don't expect this program to run in linear time on graphs of bounded degree, but we use this as another example, comparing the runtime performance of the original and new compiler implementations. We include grids as an example of an input that is not Series-Parallel.

\vspace{0.2em}
\begin{figure}[H]
\centering
\noindent
\scalebox{.73}{\fbox{\begin{minipage}{17.5cm}
\begin{allintypewriter}
Main = \{par, seq\}!; del; if node then fail

\medskip
\setlength{\tabcolsep}{16pt}
\vspace{2.5mm}
\begin{tabular}{  p{6.0cm}  p{7.8cm}  }

    \vspace{-1mm} par(a,b,x,y:list) & \vspace{-2mm} seq(a,b,x,y,z:list) \\

    \vspace{-2mm}
    \adjustbox{valign=t}{\begin{tikzpicture}[every node/.style={inner sep=0pt, text width=6.5mm, align=center}]
        \node (a) at (0.0,0)     [draw,circle,thick] {x};
        \node (b) at (1.5,0)     [draw,circle,thick] {y};

        \node (c) at (2.5,0)     {$\Rightarrow$};

        \node (d) at (3.5,0)     [draw,circle,thick] {x};
        \node (e) at (5.0,0)     [draw,circle,thick] {y};

        \node (A) at (0.0,-.52) {\tiny{1}};
        \node (B) at (1.5,-.52) {\tiny{2}};
        \node (D) at (3.5,-.52) {\tiny{1}};
        \node (E) at (5.0,-.52) {\tiny{2}};

        \draw (a) edge[->,thick,bend left=25] node[above, yshift=2.5pt] {a} (b)
              (a) edge[->,thick,bend left=-25] node[above, yshift=2.5pt] {b} (b)
              (d) edge[->,thick] node[above, yshift=2.5pt] {a} (e);
    \end{tikzpicture}}

    &

    \vspace{-2mm}
    \adjustbox{valign=t}{\begin{tikzpicture}[every node/.style={inner sep=0pt, text width=6.5mm, align=center}]
        \node (a) at (0.0,0)     [draw,circle,thick] {x};
        \node (b) at (1.5,0)     [draw,circle,thick] {y};
        \node (c) at (3.0,0)     [draw,circle,thick] {z};

        \node (d) at (4.0,0)     {$\Rightarrow$};

        \node (e) at (5.0,0)     [draw,circle,thick] {x};
        \node (f) at (6.5,0)     [draw,circle,thick] {z};

        \node (A) at (0.0,-.52) {\tiny{1}};
        \node (C) at (3.0,-.52) {\tiny{2}};
        \node (E) at (5.0,-.52) {\tiny{1}};
        \node (F) at (6.5,-.52) {\tiny{2}};

        \draw (a) edge[->,thick] node[above, yshift=2.5pt] {a} (b)
              (b) edge[->,thick] node[above, yshift=2.5pt] {b} (c)
              (e) edge[->,thick] node[above, yshift=2.5pt] {a} (f);
    \end{tikzpicture}}
    \\
\end{tabular}
\begin{tabular}{  p{6.0cm}  p{7.8cm}  }

    \vspace{-1mm} del(a,x,y:list) & \vspace{-2mm} node(x:list) \\

    \vspace{-2mm}
    \adjustbox{valign=t}{\begin{tikzpicture}[every node/.style={inner sep=0pt, text width=6.5mm, align=center}]
        \node (a) at (0.0,0) [draw,circle,thick] {x};
        \node (b) at (1.5,0) [draw,circle,thick] {y};

        \node (c) at (2.5,0) {$\Rightarrow$};

        \node (d) at (3.5,0) {$\emptyset$};

        \draw (a) edge[->,thick] node[above, yshift=2.5pt] {a} (b);
    \end{tikzpicture}}

    &

    \vspace{-2mm}
    \adjustbox{valign=t}{\begin{tikzpicture}[every node/.style={inner sep=0pt, text width=6.5mm, align=center}]
        \node (a) at (0.0,0) [draw,circle,thick] {x};

        \node (b) at (1.0,0) {$\Rightarrow$};

        \node (c) at (2.0,0) [draw,circle,thick] {x};

        \node (A) at (0.0,-.52) {\tiny{1}};
        \node (C) at (2.0,-.52) {\tiny{1}};
    \end{tikzpicture}}
    \\
\end{tabular}
\end{allintypewriter}
\end{minipage}
}}
\caption{GP\,2 Program \texttt{is-series-par.gp2}}
\label{fig:is-series-par-gp2}
\end{figure}

\vspace{0.4em}
\begin{figure}[H]
\begin{subfigure}{.5\textwidth}
    \centering
    \begin{tikzpicture}[scale=0.7]
\begin{axis}[
xlabel={Number of nodes in input},
ylabel={Execution time (ms)},
xmin=0, xmax=70000,
ymin=0, ymax=20000,
legend pos=north west,
ymajorgrids=true,
grid style=dashed,
yticklabel style={/pgf/number format/fixed},
]
\addplot[color=performanceYellow, mark=square*] 
coordinates {
    (8000,192.23)
    (16000,686.13)
    (24000,1510.51)
    (32000,2701.79)
    (40000,4523.52)
    (48000,7183.72)
    (56000,9598.38)
    (64000,12971.47)
};
\addplot[color=performanceBlue, mark=square*] 
coordinates {
    (8000,208.05)
    (16000,809.62)
    (24000,2000.83)
    (32000,3835.29)
    (40000,6452.18)
    (48000,9655.86)
    (56000,13657.45)
    (64000,18329.79)
};
\addlegendentry{Old Impl.}
\addlegendentry{New Impl.}
\end{axis}
\end{tikzpicture}
    \caption{List Reduction}
\end{subfigure}
\begin{subfigure}{.5\textwidth}
    \centering
    \begin{tikzpicture}[scale=0.7]
\begin{axis}[
xlabel={Number of nodes in input},
ylabel={Execution time (ms)},
xmin=0, xmax=1100000,
ymin=0, ymax=2400,
legend pos=north west,
ymajorgrids=true,
grid style=dashed,
yticklabel style={/pgf/number format/fixed},
]
\addplot[color=performanceYellow, mark=square*] 
coordinates {
    (90000,245.57)
    (160000,352.80)
    (250000,475.35)
    (360000,630.61)
    (490000,846.89)
    (640000,1105.30)
    (810000,1387.21)
    (1000000,1717.52)
};
\addplot[color=performanceBlue, mark=square*] 
coordinates {
    (90000,239.92)
    (160000,395.26)
    (250000,593.71)
    (360000,805.09)
    (490000,1087.21)
    (640000,1415.69)
    (810000,1785.22)
    (1000000,2202.54)
};
\addlegendentry{Old Impl.}
\addlegendentry{New Impl.}
\end{axis}
\end{tikzpicture}
    \caption{Grid Reduction}
\end{subfigure}
\vspace{0.8em}
\caption{Measured Performance of \texttt{is-series-par.gp2}}
\label{fig:is-series-par-timing}
\end{figure}


\section{Generation Performance}

\vspace{-0.5em}
\noindent
\begin{figure}[H]
\begin{subfigure}{.24\textwidth}
    \centering
    \begin{tikzpicture}[scale=0.68]
    \node (a) at (-1.500,1.333)  [draw,circle,thick] {\,};
    \node (b) at (0.000,1.333)   [draw,circle,thick] {\,};
    \node (c) at (1.500,1.333)   [draw,circle,thick] {\,};
    \node (d) at (-1.500,0.000)  [draw,circle,thick] {\,};
    \node (e) at (0.000,0.000)   [draw,circle,thick] {\,};
    \node (f) at (1.500,0.000)   [draw,circle,thick] {\,};
    \node (g) at (-1.500,-1.333) [draw,circle,thick] {\,};
    \node (h) at (0.000,-1.333)  [draw,circle,thick] {\,};
    \node (i) at (1.500,-1.333)  [draw,circle,thick] {\,};
\end{tikzpicture}
    \vspace{0.2em}
    \caption{Discrete Graph}
\end{subfigure}
\begin{subfigure}{.25\textwidth}
    \centering
    \begin{tikzpicture}[scale=0.68]
    \node (a) at (0.000,1.333)   [draw,circle,thick] {\,};
    \node (b) at (1.000,0.000)   [draw,circle,thick] {\,};
    \node (c) at (-1.000,0.000)  [draw,circle,thick] {\,};
    \node (d) at (1.500,-1.333)  [draw,circle,thick] {\,};
    \node (e) at (0.500,-1.333)  [draw,circle,thick] {\,};
    \node (f) at (-0.500,-1.333) [draw,circle,thick] {\,};
    \node (g) at (-1.500,-1.333) [draw,circle,thick] {\,};
    
    \draw (a) edge[->, thick] (b)
          (a) edge[->, thick] (c)
          (b) edge[->, thick] (d)
          (b) edge[->, thick] (e)
          (c) edge[->, thick] (f)
          (c) edge[->, thick] (g);
\end{tikzpicture}
    \vspace{0.2em}
    \caption{Binary Tree}
\end{subfigure}
\begin{subfigure}{.25\textwidth}
    \centering
    \begin{tikzpicture}[scale=0.68]
    \node (a) at (0.000,0.000)   [draw,circle,thick] {\,};
    \node (b) at (0.000,1.333)   [draw,circle,thick] {\,};
    \node (c) at (0.943,0.943)   [draw,circle,thick] {\,};
    \node (d) at (1.333,0.000)   [draw,circle,thick] {\,};
    \node (e) at (0.943,-0.943)  [draw,circle,thick] {\,};
    \node (f) at (0.000,-1.333)  [draw,circle,thick] {\,};
    \node (g) at (-0.943,-0.943) [draw,circle,thick] {\,};
    \node (h) at (-1.333,0.000)  [draw,circle,thick] {\,};
    \node (i) at (-0.943,0.943)  [draw,circle,thick] {\,};
    
    \draw (a) edge[->, thick] (b)
          (c) edge[->, thick] (a)
          (a) edge[->, thick] (d)
          (e) edge[->, thick] (a)
          (a) edge[->, thick] (f)
          (g) edge[->, thick] (a)
          (a) edge[->, thick] (h)
          (i) edge[->, thick] (a);
\end{tikzpicture}
    \vspace{0.2em}
    \caption{Star Graph}
\end{subfigure}
\begin{subfigure}{.24\textwidth}
    \centering
    \begin{tikzpicture}[scale=0.68]
    \node (a) at (0.000,1.333)   [draw,circle,thick] {\,};
    \node (b) at (-0.750,0.000)  [draw,circle,thick] {\,};
    \node (c) at (0.750,0.000)   [draw,circle,thick] {\,};
    \node (d) at (-1.500,-1.333) [draw,circle,thick] {\,};
    \node (e) at (0.000,-1.333)  [draw,circle,thick] {\,};
    \node (f) at (1.500,-1.333)  [draw,circle,thick] {\,};
    
    \draw (a) edge[->, thick] (b)
          (a) edge[->, thick] (c)
          (b) edge[->, thick] (d)
          (b) edge[->, thick] (e)
          (c) edge[->, thick] (b)
          (c) edge[->, thick] (e)
          (c) edge[->, thick] (f)
          (e) edge[->, thick] (d)
          (f) edge[->, thick] (e);
\end{tikzpicture}
    \vspace{0.2em}
    \caption{Sierpinski Graph}
\end{subfigure}
\vspace{0.8em}
\caption{Generated Graph Classes}
\label{fig:graph-types}
\end{figure}

We don't expect any complexity improvement for \enquote{generation} programs, however we include this class of programs to allow us to verify that this is the case. Our first test case is the generation of discrete graphs, our second is full binary trees, our third is \enquote{star graphs}, and our final is Sierpinski graphs using Plump's program, originally written for GP\,1 \cite{Taentzer-Biermann-Bisztray-Boneva-Boronat-Geiger-Geiss-Horvath-Kniemeyer-Mens-Ness-Plump-Vajk08a}.

\vspace{0.2em}
\begin{figure}[H]
\centering
\noindent
\scalebox{.73}{\fbox{\begin{minipage}{17.5cm}
\begin{allintypewriter}
Main = try init then (gen!; del); finish

\medskip
\setlength{\tabcolsep}{16pt}
\vspace{2.5mm}
\begin{tabular}{  p{6.9cm}  p{6.9cm}  }

    \vspace{-1mm} init(n:int) & \vspace{-2mm} gen(n,m:int) \\

    \vspace{-2mm}
    \adjustbox{valign=t}{\begin{tikzpicture}[every node/.style={inner sep=0pt, text width=6.5mm, align=center}]
        \node (a) at (0.0,0)     [draw,circle,thick,double,double distance=0.4mm] {n};

        \node (b) at (1.0,0)     {$\Rightarrow$};

        \node (c) at (2.0,0)     [draw,circle,thick,double,double distance=0.4mm] {n};
        \node (d) at (3.5,0)     [draw,circle,thick,double,double distance=0.4mm] {\scriptsize{n-1}};

        \node (A) at (0.0,-.52) {\tiny{1}};
        \node (C) at (2.0,-.52) {\tiny{1}};

        \draw (c) edge[->,thick] node[above, yshift=2.5pt] {\,} (d);
    \end{tikzpicture}}

    \vspace{1mm} where n > 1
    \vspace{5mm}

    &

    \vspace{-2mm}
    \adjustbox{valign=t}{\begin{tikzpicture}[every node/.style={inner sep=0pt, text width=6.5mm, align=center}]
        \node (a) at (0.0,0)     [draw,circle,thick,double,double distance=0.4mm] {n};
        \node (b) at (1.5,0)     [draw,circle,thick,double,double distance=0.4mm] {m};

        \node (c) at (2.5,0)     {$\Rightarrow$};

        \node (d) at (3.5,0)     [draw,circle,thick] {n};
        \node (e) at (5.0,0)     [draw,circle,thick,double,double distance=0.4mm] {m};
        \node (f) at (6.5,0)     [draw,circle,thick,double,double distance=0.4mm] {\scriptsize{m-1}};

        \node (A) at (0.0,-.52) {\tiny{1}};
        \node (B) at (1.5,-.52) {\tiny{2}};
        \node (D) at (3.5,-.52) {\tiny{1}};
        \node (E) at (5.0,-.52) {\tiny{2}};

        \draw (a) edge[->,thick] node[above, yshift=2.5pt] {\,} (b)
              (e) edge[->,thick] node[above, yshift=2.5pt] {\,} (f);
    \end{tikzpicture}}

    \vspace{1mm} where m > 1
    \vspace{5mm} \\
\end{tabular}

\begin{tabular}{  p{6.9cm}  p{6.9cm}  }

    \vspace{-1mm} del(n,m:int) & \vspace{-2mm} finish() \\

    \vspace{-2mm}
    \adjustbox{valign=t}{\begin{tikzpicture}[every node/.style={inner sep=0pt, text width=6.5mm, align=center}]
        \node (a) at (0.0,0)     [draw,circle,thick,double,double distance=0.4mm] {n};
        \node (b) at (1.5,0)     [draw,circle,thick,double,double distance=0.4mm] {m};

        \node (c) at (2.5,0)     {$\Rightarrow$};

        \node (d) at (3.5,0)     [draw,circle,thick] {n};
        \node (e) at (5.0,0)     [draw,circle,thick,double,double distance=0.4mm] {m};

        \node (A) at (0.0,-.52) {\tiny{1}};
        \node (B) at (1.5,-.52) {\tiny{2}};
        \node (D) at (3.5,-.52) {\tiny{1}};
        \node (E) at (5.0,-.52) {\tiny{2}};

        \draw (a) edge[->,thick] node[above, yshift=2.5pt] {\,} (b);
    \end{tikzpicture}}

    &

    \vspace{-2mm}
    \adjustbox{valign=t}{\begin{tikzpicture}[every node/.style={inner sep=0pt, text width=6.5mm, align=center}]
        \node (a) at (0.0,0)     [draw,circle,thick,double,double distance=0.4mm] {1};

        \node (b) at (1.0,0)     {$\Rightarrow$};

        \node (c) at (2.0,0)     [draw,circle,thick] {1};

        \node (A) at (0.0,-.52) {\tiny{1}};
        \node (C) at (2.0,-.52) {\tiny{1}};
    \end{tikzpicture}}
    \\
\end{tabular}
\end{allintypewriter}
\end{minipage}
}}
\caption{GP\,2 Program \texttt{gen-discrete.gp2}}
\label{fig:gen-discrete-gp2}
\end{figure}

\vspace{0.2em}
\begin{figure}[H]
\centering
\noindent
\scalebox{.73}{\fbox{\begin{minipage}{17.5cm}
\begin{allintypewriter}
Main = init; (gen!; ret; step!)!; finish

\medskip
\setlength{\tabcolsep}{16pt}
\vspace{2.5mm}
\begin{tabular}{  p{4.5cm}  p{4.7cm}  p{4.6cm}  }

    \vspace{-1mm} init(n:int) & \vspace{-2mm} ret(i,j:int) & \vspace{-2mm} finish(n:int) \\

    \vspace{-2mm}
    \adjustbox{valign=t}{\begin{tikzpicture}[every node/.style={inner sep=0pt, text width=6.5mm, align=center}]
        \node (a) at (0.0,0)     [draw,circle,thick,double,double distance=0.4mm] {n};

        \node (b) at (1.0,0)     {$\Rightarrow$};

        \node (c) at (2.0,0)     [draw,circle,fill=gp2grey,thick,double,double distance=0.4mm] {n};
        \node (d) at (3.5,0)     [draw,circle,thick,double,double distance=0.4mm] {0};

        \node (A) at (0.0,-0.52) {\tiny{1}};
        \node (C) at (3.5,-0.52) {\tiny{1}};
    \end{tikzpicture}}

    &

    \vspace{-2mm}
    \adjustbox{valign=t}{\begin{tikzpicture}[every node/.style={inner sep=0pt, text width=6.5mm, align=center}]
        \node (a) at (0.0,-1.5)  [draw,circle,thick,double,double distance=0.4mm] {j};
        \node (b) at (0.8,0)     [draw,circle,thick] {i};

        \node (c) at (1.8,-0.75) {$\Rightarrow$};

        \node (d) at (2.8,-1.5)  [draw,circle,thick] {j};
        \node (e) at (3.6,0)     [draw,circle,thick,double,double distance=0.4mm] {i};

        \node (A) at (0.0,-2.02) {\tiny{2}};
        \node (C) at (0.8,-0.52) {\tiny{1}};
        \node (E) at (2.8,-2.02) {\tiny{2}};
        \node (F) at (3.6,-0.52) {\tiny{1}};

        \draw (b) edge[->,thick] node[above, yshift=2.5pt] {\,} (a)
              (e) edge[->,thick] node[above, yshift=2.5pt] {\,} (d);
    \end{tikzpicture}}

    &

    \vspace{-2mm}
    \adjustbox{valign=t}{\begin{tikzpicture}[every node/.style={inner sep=0pt, text width=6.5mm, align=center}]
        \node (a) at (0.0,0)     [draw,circle,fill=gp2grey,thick,double,double distance=0.4mm] {n};
        \node (b) at (1.5,0)     [draw,circle,thick,double,double distance=0.4mm] {0};

        \node (c) at (2.5,0)     {$\Rightarrow$};

        \node (d) at (3.5,0)     [draw,circle,thick] {n};

        \node (B) at (1.5,-0.52) {\tiny{1}};
        \node (D) at (3.5,-0.52) {\tiny{1}};
    \end{tikzpicture}}
    \\
\end{tabular}
\begin{tabular}{  p{7.6cm}  p{6.2cm} }

    \vspace{-1mm} gen(i,n:init) & \vspace{-2mm} step(i,j,n:int) \\

    \vspace{-2mm}
    \adjustbox{valign=t}{\begin{tikzpicture}[every node/.style={inner sep=0pt, text width=6.5mm, align=center}]
        \node (a) at (0.0,0)     [draw,circle,fill=gp2grey,thick,double,double distance=0.4mm] {n};
        \node (b) at (1.5,0)     [draw,circle,thick,double,double distance=0.4mm] {i};

        \node (c) at (2.5,-0.75) {$\Rightarrow$};

        \node (d) at (3.5,0)     [draw,circle,thick,double,double distance=0.4mm] {n};
        \node (e) at (5.0,0)     [draw,circle,thick] {i};
        \node (f) at (4.2,-1.5)  [draw,circle,thick,double,double distance=0.4mm] {\scriptsize{i+1}};
        \node (g) at (5.8,-1.5)  [draw,circle,thick] {\scriptsize{i+1}};

        \node (A) at (0.0,-0.52) {\tiny{1}};
        \node (C) at (1.5,-0.52) {\tiny{2}};
        \node (E) at (3.5,-0.52) {\tiny{1}};
        \node (F) at (5.0,-0.52) {\tiny{2}};

        \draw (e) edge[->,thick] node[above, yshift=2.5pt] {\,} (f)
              (e) edge[->,thick] node[above, yshift=2.5pt] {\,} (g);
    \end{tikzpicture}}

    \vspace{1mm} where i < n and outdeg(2) = 0
    \vspace{5mm} 

    &

    \vspace{-2mm}
    \adjustbox{valign=t}{\begin{tikzpicture}[every node/.style={inner sep=0pt, text width=6.5mm, align=center}]
        \node (a) at (0.0,0)     [draw,circle,thick,fill=gp2grey,double,double distance=0.4mm] {n};
        \node (b) at (1.5,0)     [draw,circle,thick] {i};
        \node (c) at (0.7,-1.5)  [draw,circle,thick,double,double distance=0.4mm] {\scriptsize{i+1}};

        \node (d) at (2.5,-0.75) {$\Rightarrow$};

        \node (e) at (3.5,0)     [draw,circle,thick,fill=gp2grey,double,double distance=0.4mm] {n};
        \node (f) at (5.0,0)     [draw,circle,thick] {i};
        \node (g) at (4.2,-1.5)  [draw,circle,thick,double,double distance=0.4mm] {\scriptsize{i+1}};

        \node (A) at (0.0,-0.52) {\tiny{2}};
        \node (C) at (1.5,-0.52) {\tiny{1}};
        \node (E) at (3.5,-0.52) {\tiny{2}};
        \node (F) at (5.0,-0.52) {\tiny{1}};

        \draw (b) edge[->,thick] node[above, yshift=2.5pt] {\,} (c)
              (f) edge[->,thick] node[above, yshift=2.5pt] {\,} (g);
    \end{tikzpicture}}

    \vspace{1mm} where j < n
    \vspace{5mm}
    \\
\end{tabular}
\end{allintypewriter}
\end{minipage}
}}
\caption{GP\,2 Program \texttt{gen-tree.gp2}}
\label{fig:gen-tree-gp2}
\end{figure}

\vspace{0.2em}
\begin{figure}[H]
\centering
\noindent
\scalebox{.73}{\fbox{\begin{minipage}{17.5cm}
\begin{allintypewriter}
Main = \{gen1, gen2\}!; \{fin1, fin2\}

\medskip
\setlength{\tabcolsep}{16pt}
\vspace{2.5mm}
\begin{tabular}{  p{6.9cm}  p{6.9cm}  }

    \vspace{-1mm} gen1(n:int) & \vspace{-2mm} gen2(n:int) \\

    \vspace{-2mm}
    \adjustbox{valign=t}{\begin{tikzpicture}[every node/.style={inner sep=0pt, text width=6.5mm, align=center}]
        \node (a) at (0.0,0)     [draw,circle,thick,double,double distance=0.4mm] {n};

        \node (c) at (1.0,0)     {$\Rightarrow$};

        \node (d) at (2.0,0)     [draw,circle,fill=gp2grey,thick,double,double distance=0.4mm] {\scriptsize{n-1}};
        \node (e) at (3.5,0)     [draw,circle,thick] {n};

        \node (A) at (0.0,-.52) {\tiny{1}};
        \node (D) at (2.0,-.52) {\tiny{1}};

        \draw (d) edge[->,thick] node[above, yshift=2.5pt] {n} (e);
    \end{tikzpicture}}

    \vspace{1mm} where n > 1
    \vspace{5mm}

    &

    \vspace{-2mm}
    \adjustbox{valign=t}{\begin{tikzpicture}[every node/.style={inner sep=0pt, text width=6.5mm, align=center}]
        \node (a) at (0.0,0)     [draw,circle,fill=gp2grey,thick,double,double distance=0.4mm] {n};

        \node (c) at (1.0,0)     {$\Rightarrow$};

        \node (d) at (2.0,0)     [draw,circle,thick,double,double distance=0.4mm] {\scriptsize{n-1}};
        \node (e) at (3.5,0)     [draw,circle,thick] {n};

        \node (A) at (0.0,-.52) {\tiny{1}};
        \node (D) at (2.0,-.52) {\tiny{1}};

        \draw (d) edge[->,thick] node[above, yshift=2.5pt] {n} (e);
    \end{tikzpicture}}

    \vspace{1mm} where n > 1
    \vspace{5mm} \\
\end{tabular}

\begin{tabular}{  p{6.9cm}  p{6.9cm}  }

    \vspace{-1mm} fin1() & \vspace{-2mm} fin2() \\

    \vspace{-2mm}
    \adjustbox{valign=t}{\begin{tikzpicture}[every node/.style={inner sep=0pt, text width=6.5mm, align=center}]
        \node (a) at (0.0,0) [draw,circle,thick,double,double distance=0.4mm] {1};

        \node (b) at (1.0,0) {$\Rightarrow$};

        \node (c) at (2.0,0) [draw,circle,thick] {1};

        \node (A) at (0.0,-.52) {\tiny{1}};
        \node (C) at (2.0,-.52) {\tiny{1}};
    \end{tikzpicture}}

    &

    \vspace{-2mm}
    \adjustbox{valign=t}{\begin{tikzpicture}[every node/.style={inner sep=0pt, text width=6.5mm, align=center}]
        \node (a) at (0.0,0) [draw,circle,fill=gp2grey,thick,double,double distance=0.4mm] {1};

        \node (b) at (1.0,0) {$\Rightarrow$};

        \node (c) at (2.0,0) [draw,circle,thick] {1};

        \node (A) at (0.0,-.52) {\tiny{1}};
        \node (C) at (2.0,-.52) {\tiny{1}};
    \end{tikzpicture}}
    \\
\end{tabular}
\end{allintypewriter}
\end{minipage}
}}
\caption{GP\,2 Program \texttt{gen-star.gp2}}
\label{fig:gen-star-gp2}
\end{figure}

\vspace{0.2em}
\begin{figure}[H]
\centering
\noindent
\scalebox{.73}{\fbox{\begin{minipage}{17.5cm}
\begin{allintypewriter}
Main = init; (inc; expand!)!; cleanup

\medskip
\setlength{\tabcolsep}{16pt}
\vspace{2.5mm}
\begin{tabular}{  p{8.0cm}  p{5.8cm}  }

    \vspace{-1mm} init(m:int) & \vspace{-2mm} inc(m,n:int) \\

    \vspace{-2mm}
    \adjustbox{valign=t}{\begin{tikzpicture}[every node/.style={inner sep=0pt, text width=6.5mm, align=center}]
        \node (a) at (0.0,-0)    [draw,circle,thick,double,double distance=0.4mm] {m};

        \node (b) at (1.0,-0.75) {$\Rightarrow$};

        \node (c) at (2.0,0)     [draw,circle,thick,double,double distance=0.4mm] {\ssmall{m:0}};
        \node (d) at (3.5,0)     [draw,circle,thick] {1};
        \node (e) at (2.75,-1.5) [draw,circle,thick] {0};
        \node (f) at (4.25,-1.5) [draw,circle,thick] {0};

        \node (A) at (0.0,-0.52) {\tiny{1}};
        \node (C) at (2.0,-0.52) {\tiny{1}};

        \draw (d) edge[->,thick] node[above, xshift=-3pt] {1} (e)
              (d) edge[->,thick] node[above, xshift=3pt] {0} (f)
              (f) edge[->,thick] node[above, yshift=2.5pt] {2} (e);
    \end{tikzpicture}}

    &

    \vspace{-2mm}
    \adjustbox{valign=t}{\begin{tikzpicture}[every node/.style={inner sep=0pt, text width=6.5mm, align=center}]
        \node (a) at (0.0,0)     [draw,circle,thick,double,double distance=0.4mm] {\ssmall{m:n}};

        \node (b) at (1.0,0)     {$\Rightarrow$};

        \node (c) at (2.0,0)     [draw,circle,thick,double,double distance=0.4mm] {\fontsize{5}{5}\selectfont{m:n+1}};

        \node (A) at (0.0,-0.52) {\tiny{1}};
        \node (C) at (2.0,-0.52) {\tiny{1}};
    \end{tikzpicture}}

    \vspace{1mm} where m > n
    \vspace{5mm} \\
\end{tabular}
\begin{tabular}{  p{8.0cm}  p{5.8cm}  }

    \vspace{-1mm} expand(m,n,p,q:int) & \vspace{-2mm} cleanup(m,n:int) \\

    \vspace{-2mm}
    \adjustbox{valign=t}{\begin{tikzpicture}[every node/.style={inner sep=0pt, text width=6.5mm, align=center}]
        \node (a) at (0.0,0)      [draw,circle,thick,double,double distance=0.4mm] {\ssmall{m:n}};
        \node (b) at (1.5,0)      [draw,circle,thick] {n};
        \node (c) at (0.75,-1.5)  [draw,circle,thick] {q};
        \node (d) at (2.25,-1.5)  [draw,circle,thick] {p};

        \node (e) at (3.75,-0.75) {$\Rightarrow$};

        \node (f) at (4.25,0)     [draw,circle,thick,double,double distance=0.4mm] {\ssmall{m:n}};
        \node (g) at (5.75,0)     [draw,circle,thick] {\scriptsize{n+1}};
        \node (h) at (5.0,-1.5)   [draw,circle,thick] {\scriptsize{n+1}};
        \node (i) at (6.5,-1.5)   [draw,circle,thick] {\scriptsize{n+1}};
        \node (j) at (4.25,-3.0)  [draw,circle,thick] {q};
        \node (k) at (5.75,-3.0)  [draw,circle,thick] {0};
        \node (l) at (7.25,-3.0)  [draw,circle,thick] {p};

        \node (A) at (0.0,-0.52)  {\tiny{1}};
        \node (B) at (1.5,-0.52)  {\tiny{2}};
        \node (C) at (0.75,-2.02) {\tiny{4}};
        \node (D) at (2.25,-2.02) {\tiny{3}};
        \node (F) at (4.25,-0.52) {\tiny{1}};
        \node (G) at (5.75,-0.52) {\tiny{2}};
        \node (H) at (4.25,-3.52) {\tiny{4}};
        \node (I) at (7.25,-3.52) {\tiny{3}};

        \draw (b) edge[->,thick] node[above, xshift=3pt] {0} (d)
              (b) edge[->,thick] node[above, xshift=-3pt] {1} (c)
              (d) edge[->,thick] node[above, yshift=2.5pt] {2} (c)
              (g) edge[->,thick] node[above, xshift=3pt] {0} (i)
              (g) edge[->,thick] node[above, xshift=-3pt] {1} (h)
              (i) edge[->,thick] node[above, yshift=2.5pt] {2} (h)
              (h) edge[->,thick] node[above, xshift=3pt] {0} (k)
              (h) edge[->,thick] node[above, xshift=-3pt] {1} (j)
              (k) edge[->,thick] node[above, yshift=2.5pt] {2} (j)
              (i) edge[->,thick] node[above, xshift=3pt] {0} (l)
              (i) edge[->,thick] node[above, xshift=-3pt] {1} (k)
              (l) edge[->,thick] node[above, yshift=2.5pt] {2} (k);
    \end{tikzpicture}}
    \vspace{6mm}

    &

    \vspace{-2mm}
    \adjustbox{valign=t}{\begin{tikzpicture}[every node/.style={inner sep=0pt, text width=6.5mm, align=center}]
        \node (a) at (0.0,0)     [draw,circle,thick,double,double distance=0.4mm] {\ssmall{m:n}};

        \node (b) at (1.0,0)     {$\Rightarrow$};

        \node (c) at (2.0,0)     {$\emptyset$};
    \end{tikzpicture}}
    \\
\end{tabular}
\end{allintypewriter}
\end{minipage}
}}
\caption{GP\,2 Program \texttt{gen-sierpinski.gp2}}
\label{fig:gen-sierpinski-gp2}
\end{figure}

\vspace{0.4em}
\begin{figure}[H]
\begin{subfigure}{.5\textwidth}
    \centering
    \begin{tikzpicture}[scale=0.7]
\begin{axis}[
xlabel={Requested number of nodes},
ylabel={Execution time (ms)},
xmin=0, xmax=2200000,
ymin=0, ymax=2000,
legend pos=north west,
ymajorgrids=true,
grid style=dashed,
yticklabel style={/pgf/number format/fixed},
]
\addplot[color=performanceYellow, mark=square*] 
coordinates {
    (200000,176.33)
    (400000,310.99)
    (600000,461.95)
    (800000,598.24)
    (1000000,756.75)
    (1200000,931.06)
    (1400000,1091.21)
    (1600000,1282.00)
    (1800000,1469.81)
    (2000000,1698.61)
};
\addplot[color=performanceBlue, mark=square*] 
coordinates {
    (200000,180.92)
    (400000,330.38)
    (600000,475.31)
    (800000,636.41)
    (1000000,795.72)
    (1200000,961.30)
    (1400000,1148.59)
    (1600000,1346.37)
    (1800000,1534.67)
    (2000000,1754.75)
};
\addlegendentry{Old Impl.}
\addlegendentry{New Impl.}
\end{axis}
\end{tikzpicture}
    \caption{\texttt{gen-discrete.gp2} Performance}
\end{subfigure}
\begin{subfigure}{.5\textwidth}
    \centering
    \begin{tikzpicture}[scale=0.7]
\begin{axis}[
xlabel={Requested tree depth},
ylabel={Execution time (ms)},
xmin=12, xmax=19,
ymin=0, ymax=750,
legend pos=north west,
ymajorgrids=true,
grid style=dashed,
yticklabel style={/pgf/number format/fixed},
]
\addplot[color=performanceYellow, mark=square*] 
coordinates {
    (13,52.65)
    (14,59.56)
    (15,99.05)
    (16,186.83)
    (17,293.73)
    (18,599.64)
};
\addplot[color=performanceBlue, mark=square*] 
coordinates {
    (13,43.47)
    (14,78.21)
    (15,106.24)
    (16,191.48)
    (17,344.52)
    (18,681.29)
};
\addlegendentry{Old Impl.}
\addlegendentry{New Impl.}
\end{axis}
\end{tikzpicture}
    \caption{\texttt{gen-tree.gp2} Performance}
\end{subfigure}
\begin{subfigure}{.5\textwidth}
    \centering
    \vspace{1.8em}
    \begin{tikzpicture}[scale=0.7]
\begin{axis}[
xlabel={Requested number of nodes},
ylabel={Execution time (ms)},
xmin=0, xmax=1125000,
ymin=0, ymax=1600,
legend pos=north west,
ymajorgrids=true,
grid style=dashed,
yticklabel style={/pgf/number format/fixed},
]
\addplot[color=performanceYellow, mark=square*] 
coordinates {
    (100000,151.20)
    (200000,289.92)
    (300000,394.47)
    (400000,509.31)
    (500000,650.33)
    (600000,798.18)
    (700000,905.19)
    (800000,1068.03)
    (900000,1182.93)
    (1000000,1340.69)
};
\addplot[color=performanceBlue, mark=square*] 
coordinates {
    (100000,163.21)
    (200000,292.68)
    (300000,430.37)
    (400000,571.25)
    (500000,718.22)
    (600000,844.32)
    (700000,987.49)
    (800000,1129.30)
    (900000,1300.17)
    (1000000,1412.95)
};
\addlegendentry{Old Impl.}
\addlegendentry{New Impl.}
\end{axis}
\end{tikzpicture}
    \caption{\texttt{gen-star.gp2} Performance}
\end{subfigure}
\begin{subfigure}{.5\textwidth}
    \centering
    \vspace{1.8em}
    \begin{tikzpicture}[scale=0.7]
\begin{axis}[
xlabel={Requested level of recursion},
ylabel={Execution time (ms)},
xmin=5, xmax=11,
ymin=0, ymax=10000,
legend pos=north west,
ymajorgrids=true,
grid style=dashed,
yticklabel style={/pgf/number format/fixed},
]
\addplot[color=performanceYellow, mark=square*] 
coordinates {
    (6,16.77)
    (7,22.83)
    (8,54.55)
    (9,275.28)
    (10,2084.26)
};
\addplot[color=performanceBlue, mark=square*] 
coordinates {
    (6,15.74)
    (7,25.82)
    (8,97.10)
    (9,821.77)
    (10,9603.32)
};
\addlegendentry{Old Impl.}
\addlegendentry{New Impl.}
\end{axis}
\end{tikzpicture}
    \caption{\texttt{gen-sierpinski.gp2} Performance}
\end{subfigure}
\vspace{0.8em}
\caption{Measured Generation Performance}
\label{fig:generation-timing}
\end{figure}


\section{Other Program Performance}

We now look at the performance of undirected DFS \cite{Bak15a} by looking at a connected graph recognition program, and also the performance of a transitive closure program. We expect the performance of both the original and new implementations to be similar for connectedness checking, with the new implementation having the edge on grid graphs due to the new implementation of edge lists. In particular, we expect to see good a runtime speedup with the transitive closure program, which is edge search intensive.

\vspace{0.2em}
\begin{figure}[H]
\centering
\noindent
\scalebox{.73}{\fbox{\begin{minipage}{17.5cm}
\begin{allintypewriter}
Main = try init then (DFS!; Check)

DFS = fwd!; try bck else break

Check = if match then fail

\medskip
\setlength{\tabcolsep}{16pt}
\vspace{2.5mm}
\begin{tabular}{  p{6.9cm}  p{6.9cm}  }

    \vspace{-1mm} init(x:list) & \vspace{-2mm} fwd(x,y,n:list) \\

    \vspace{-2mm}
    \adjustbox{valign=t}{\begin{tikzpicture}[every node/.style={inner sep=0pt, text width=6.5mm, align=center}]
        \node (a) at (0.0,0) [draw,circle,thick] {x};

        \node (b) at (1.0,0) {$\Rightarrow$};
        
        \node (c) at (2.0,0) [draw,circle,fill=gp2grey,thick,double,double distance=0.4mm] {x};
        
        \node (A) at (0.0,-.52) {\tiny{1}};
        \node (C) at (2.0,-.52) {\tiny{1}};
    \end{tikzpicture}}

    &

    \vspace{-2mm}
    \adjustbox{valign=t}{\begin{tikzpicture}[every node/.style={inner sep=0pt, text width=6.5mm, align=center}]
        \node (a) at (0.0,0) [draw,circle,fill=gp2grey,thick,double,double distance=0.4mm] {x};
        \node (b) at (1.5,0) [draw,circle,thick] {y};

        \node (c) at (2.5,0) {$\Rightarrow$};

        \node (d) at (3.5,0) [draw,circle,fill=gp2grey,thick] {x};
        \node (e) at (5.0,0) [draw,circle,fill=gp2grey,thick,double,double distance=0.4mm] {y};

        \node (A) at (0.0,-.52) {\tiny{1}};
        \node (C) at (1.5,-.52) {\tiny{2}};
        \node (D) at (3.5,-.52) {\tiny{1}};
        \node (E) at (5.0,-.52) {\tiny{2}};

        \draw (a) edge[-,thick] node[above, yshift=2.5pt] {n} (b)
              (d) edge[-,thick,dashed] node[above, yshift=2.5pt] {n} (e);
    \end{tikzpicture}}
    \\
\end{tabular}
\begin{tabular}{  p{6.9cm}  p{6.9cm}  }

    \vspace{-1mm} match(x,z:list) & \vspace{-2mm} bck(x,y,n:list) \\

    \vspace{-2mm}
    \adjustbox{valign=t}{\begin{tikzpicture}[every node/.style={inner sep=0pt, text width=6.5mm, align=center}]
        \node (a) at (0.0,0) [draw,circle,fill=gp2grey,thick,double,double distance=0.4mm] {x};
        \node (b) at (1.5,0) [draw,circle,thick] {z};

        \node (c) at (2.5,0) {$\Rightarrow$};

        \node (d) at (3.5,0) [draw,circle,fill=gp2grey,thick,double,double distance=0.4mm] {x};
        \node (e) at (5.0,0) [draw,circle,thick] {z};

        \node (A) at (0.0,-.52) {\tiny{1}};
        \node (C) at (1.5,-.52) {\tiny{2}};
        \node (D) at (3.5,-.52) {\tiny{1}};
        \node (E) at (5.0,-.52) {\tiny{2}};
    \end{tikzpicture}}

    &

    \vspace{-2mm}
    \adjustbox{valign=t}{\begin{tikzpicture}[every node/.style={inner sep=0pt, text width=6.5mm, align=center}]
        \node (a) at (0.0,0) [draw,circle,fill=gp2grey,thick] {x};
        \node (b) at (1.5,0) [draw,circle,fill=gp2grey,thick,double,double distance=0.4mm] {y};

        \node (c) at (2.5,0) {$\Rightarrow$};

        \node (d) at (3.5,0) [draw,circle,fill=gp2grey,thick,double,double distance=0.4mm] {x};
        \node (e) at (5.0,0) [draw,circle,fill=gp2blue,thick] {y};

        \node (A) at (0.0,-.52) {\tiny{1}};
        \node (C) at (1.5,-.52) {\tiny{2}};
        \node (D) at (3.5,-.52) {\tiny{1}};
        \node (E) at (5.0,-.52) {\tiny{2}};

        \draw (a) edge[-,thick,dashed] node[above, yshift=2.5pt] {n} (b)
              (d) edge[-,thick] node[above, yshift=2.5pt] {n} (e);
    \end{tikzpicture}}
    \\
\end{tabular}
\end{allintypewriter}
\end{minipage}
}}
\caption{GP\,2 Program \texttt{is-con.gp2}}
\label{fig:is-con-gp2}
\end{figure}

\vspace{0.4em}
\begin{figure}[H]
\begin{subfigure}{.5\textwidth}
    \centering
    \begin{tikzpicture}[scale=0.7]
\begin{axis}[
xlabel={Number of nodes in input},
ylabel={Execution time (ms)},
xmin=0, xmax=1100000,
ymin=0, ymax=1000,
legend pos=north west,
ymajorgrids=true,
grid style=dashed,
yticklabel style={/pgf/number format/fixed},
]
\addplot[color=performanceYellow, mark=square*] 
coordinates {
    (100000,101.86)
    (200000,181.15)
    (300000,240.58)
    (400000,282.28)
    (500000,358.64)
    (600000,423.81)
    (700000,493.50)
    (800000,574.86)
    (900000,630.43)
    (1000000,701.68)
};
\addplot[color=performanceBlue, mark=square*] 
coordinates {
    (100000,123.99)
    (200000,192.66)
    (300000,287.51)
    (400000,359.60)
    (500000,453.08)
    (600000,535.72)
    (700000,631.87)
    (800000,724.03)
    (900000,806.85)
    (1000000,894.51)
};
\addlegendentry{Old Impl.}
\addlegendentry{New Impl.}
\end{axis}
\end{tikzpicture}
    \caption{Performance on Discrete Graphs}
\end{subfigure}
\begin{subfigure}{.5\textwidth}
    \centering
    \begin{tikzpicture}[scale=0.7]
\begin{axis}[
xlabel={Number of nodes in input},
ylabel={Execution time (ms)},
xmin=0, xmax=1100000,
ymin=0, ymax=5000,
legend pos=north west,
ymajorgrids=true,
grid style=dashed,
yticklabel style={/pgf/number format/fixed},
]
\addplot[color=performanceYellow, mark=square*] 
coordinates {
    (90000,592.46)
    (160000,799.34)
    (250000,1256.02)
    (360000,1729.80)
    (490000,2435.70)
    (640000,3057.20)
    (810000,3773.09)
    (1000000,4667.30)
};
\addplot[color=performanceBlue, mark=square*] 
coordinates {
    (90000,386.36)
    (160000,659.67)
    (250000,998.99)
    (360000,1448.66)
    (490000,1984.27)
    (640000,2615.16)
    (810000,3332.27)
    (1000000,4084.19)
};
\addlegendentry{Old Impl.}
\addlegendentry{New Impl.}
\end{axis}
\end{tikzpicture}
    \caption{Performance on Grid Graphs}
\end{subfigure}
\vspace{0.8em}
\caption{Measured Performance of \texttt{is-con.gp2}}
\label{fig:is-con-timing}
\end{figure}

\vspace{0.2em}
\begin{figure}[H]
\centering
\noindent
\scalebox{.73}{\fbox{\begin{minipage}{17.5cm}
\begin{allintypewriter}
Main = link!

\medskip
\setlength{\tabcolsep}{16pt}
\vspace{2.5mm}
\begin{tabular}{  p{13.8cm}  }

    \vspace{-1mm} link(a,b,x,y,z:list) \\

    \vspace{-2mm}
    \adjustbox{valign=t}{\begin{tikzpicture}[every node/.style={inner sep=0pt, text width=6.5mm, align=center}]
        \node (a) at (0.0,0)     [draw,circle,thick] {x};
        \node (b) at (1.5,0)     [draw,circle,thick] {y};
        \node (c) at (3.0,0)     [draw,circle,thick] {z};

        \node (d) at (4.0,0)     {$\Rightarrow$};

        \node (e) at (5.0,0)     [draw,circle,thick] {x};
        \node (f) at (6.5,0)     [draw,circle,thick] {y};
        \node (g) at (8.0,0)     [draw,circle,thick] {z};

        \node (A) at (0.0,-.52) {\tiny{1}};
        \node (B) at (1.5,-.52) {\tiny{2}};
        \node (C) at (3.0,-.52) {\tiny{3}};
        \node (E) at (5.0,-.52) {\tiny{1}};
        \node (F) at (6.5,-.52) {\tiny{2}};
        \node (G) at (8.0,-.52) {\tiny{3}};

        \draw (a) edge[->,thick] node[above, yshift=2.5pt] {a} (b)
              (b) edge[->,thick] node[above, yshift=2.5pt] {b} (c)
              (e) edge[->,thick] node[above, yshift=2.5pt] {a} (f)
              (f) edge[->,thick] node[above, yshift=2.5pt] {b} (g)
              (e) edge[->,thick,bend left=35] node[above, yshift=2.5pt] {\,} (g);
    \end{tikzpicture}}

    \vspace{2mm} where not edge(1,3)
    \vspace{5mm} \\
\end{tabular}
\end{allintypewriter}
\end{minipage}
}}
\caption{GP\,2 Program \texttt{trans-closure.gp2}}
\label{fig:trans-closure-gp2}
\end{figure}

\vspace{0.4em}
\begin{figure}[H]
\begin{subfigure}{.5\textwidth}
    \centering
    \begin{tikzpicture}[scale=0.7]
\begin{axis}[
xlabel={Number of nodes in input},
ylabel={Execution time (ms)},
xmin=0, xmax=110,
ymin=0, ymax=42000,
legend pos=north west,
ymajorgrids=true,
grid style=dashed,
yticklabel style={/pgf/number format/fixed},
]
\addplot[color=performanceYellow, mark=square*] 
coordinates {
    (10,11.50)
    (20,16.10)
    (30,56.63)
    (40,232.39)
    (50,777.02)
    (60,2158.89)
    (70,5165.75)
    (80,11106.40)
    (90,21699.39)
    (100,39864.74)
};
\addplot[color=performanceBlue, mark=square*] 
coordinates {
    (10,10.85)
    (20,12.97)
    (30,32.32)
    (40,119.44)
    (50,401.89)
    (60,1136.29)
    (70,2781.41)
    (80,6057.80)
    (90,12036.47)
    (100,22264.00)
};
\addlegendentry{Old Impl.}
\addlegendentry{New Impl.}
\end{axis}
\end{tikzpicture}
    \caption{Performance on Linked Lists}
\end{subfigure}
\begin{subfigure}{.5\textwidth}
    \centering
    \begin{tikzpicture}[scale=0.7]
\begin{axis}[
xlabel={Number of nodes in input},
ylabel={Execution time (ms)},
xmin=0, xmax=130,
ymin=0, ymax=18000,
legend pos=north west,
ymajorgrids=true,
grid style=dashed,
yticklabel style={/pgf/number format/fixed},
]
\addplot[color=performanceYellow, mark=square*] 
coordinates {
    (25,15.46)
    (36,38.96)
    (49,145.37)
    (64,558.13)
    (81,1930.91)
    (100,6017.70)
    (121,17083.91)
};
\addplot[color=performanceBlue, mark=square*] 
coordinates {
    (25,13.79)
    (36,27.41)
    (49,96.79)
    (64,375.40)
    (81,1345.13)
    (100,4311.57)
    (121,12500.58)
};
\addlegendentry{Old Impl.}
\addlegendentry{New Impl.}
\end{axis}
\end{tikzpicture}
    \caption{Performance on Grid Graphs}
\end{subfigure}
\vspace{0.8em}
\caption{Measured Performance of \texttt{trans-closure.gp2}}
\label{fig:trans-closure-timing}
\end{figure}


\section{Results Summary}

To begin, it is evident that the new compiler vastly outperforms the old when executing the \texttt{is-discrete.gp2} program, almost appearing to be constant time in comparison. We have clearly reached linear complexity, a feat due to our node list's capacity to skip holes in the underlying node array. The new compiler also outperforms the old when running \texttt{trans-closure.gp2} by a significant constant factor. We attribute this to better cache usage; in the legacy compiler, only the indices of the first two edges are statically stored in the \lstinline{Node} type, whereas the new compiler stores several more direct pointers to nodes statically in a node's \lstinline{BigArray}. This not only means the locations of more edges are loaded into the cache in one cacheline refill, but also the loaded locations can be resolved in one memory operation rather than two, namely checking the edge array at the given index. We consider this evidence that the new compiler is superior at loading neighbouring edges efficiently.

However, the \texttt{master} compiler falls short on some test cases; the programs \texttt{gen-tree.gp2}, \texttt{gen-discrete.gp2}, \texttt{gen-star.gp2}, \texttt{gen-sierpinsk\-i.gp2}, \texttt{is-tree.gp2} and \texttt{is-series-par.g\-p2} run with a worse constant time factor and the same complexity. Many of these programs only differ by a nearly negligible factor though; the only programs that now perform substantially worse are \texttt{gen-sierpinski.gp2}, \texttt{is-series-par.gp2} and \texttt{is-tree.gp2}. It is possible that \texttt{gen-sierpinski.gp2} now performs far worse due to the fact that nodes are now accessed in the reverse order they would have been in the \texttt{legacy} compiler, making the failure somewhat artificial. Many of these failings however are likely due to excessive memory operations in the new naive linked list format when iterating through nodes.

Some programs performed better or worse depending on the type of input graph. \texttt{is-bin-dag.gp2} saw a boost in performance on linked lists, but a drop on grid graphs. On the contrary, \texttt{is-con.gp2} was found to be more performant on grid graphs but less so on linked lists. This is no consistent drop in performance, but merely a reminder that the internals of the compiler are simply different than before rather than worse.

Overall, the worst case drop in performance was by a constant factor, meaning no complexities were worsened in the observed test cases. Most programs performed similarly to before with a slightly enlarged constant, a byproduct of the memory operations a naive linked list entails.

\chapter{Conclusion} \label{chapter:conclusion}

A large number of changes were implemented in the new compiler to improve upon the old version. First and foremost, node and edge lists were added to improve iteration performance for programs with a high rate of deletion, letting generated node matching code skip large regions of deleted nodes in constant time.

On top of this, input graph parsing was fixed for graphs of arbitrary size by using Judy arrays. In the previous implementation, a user had to specify the maximum number of nodes and edges expected in input host graphs at program compile time, with the final generated code always allocating this amount of memory regardless of the actual size of input graph. In our updated implementation, this compile time knowledge is obsolete; we now dynamically grow the memory needed during parsing, allowing for arbitrary input graph size with efficient memory consumption.

Also, a method was introduced to shut down the generated program as soon as it has written the output graph without freeing any allocated heap memory properly. In general, proper management of the heap memory is advised to avoid bugs due to memory leaks or otherwise. However, one may wish to avoid doing this in the interest of runtime performance, allowing the process to terminate as quickly as possible. The newly introduced option gives the user this choice.

Finally, an option was added to require matches to both reflect and preserve root nodes. By default, GP\,2 will allow a non-root in a rule LHS to match against either a non-root or a root in a host graph, but if we insist on matches reflecting root nodes, matches can only match non-roots against non-roots. With this option employed, rooted rules are now completely reversible and finer control is gained over rule applications.


\section{Evaluation}

Overall, the new \texttt{master} compiler has been shown to maintain previous time complexities and improve on the issue of holes in some reduction programs. The \texttt{is-discrete.gp2} program has been brought down to linear complexity, making the new compiler more accurately reflect how complexity of graph programs should realistically behave. The new compiler also resolves fundamental issues with the old, such as allowing for arbitrary sizes of input graphs to programs; this makes the compiler fundamentally more correct than it was before.

However, a number of programs performed worse by a constant factor due to the overhead of pointer operations when iterating through a linked list. These issues can be avoided by not using linked lists and iterating through arrays as before, leaving it up to the user to decide how best to compile their program. In any case, no time complexity was worsened. Therefore, we deem this project a success.


\section{Future Work}

It remains future work to determine if it would make sense to add additional node lists for marked vs unmarked to allow a node search to find a marked or unmarked node in constant time. Currently, if one has a graph with arbitrarily many connected components of arbitrary size, visiting all the nodes in the graph with an undirected DFS requires quadratic time. If the DFS were to mark all the nodes in the current component, one could then imagine being able to jump to the next unvisited component in constant time, and then repeating until there are no more components. Obviously, maintaining these additional data structures requires additional time and space, and in general, will not improve the performance of programs that don't need to traverse a graph with arbitrarily many connected components.

As well as this, investigating worsened performance of programs due to linked list operations is an issue to resolve. An option is to have each linked list entry represent a range of node indices in the underlying node array rather than a single node; when iterating through nodes, a list entry may retrieve the array and iterate through it directly within its own range of nodes. When a node is deleted, the linked list entry containing it then fragments into two entries, one for the range of nodes before it and one after. This may help mitigate pointer operations and regain the speed of array iteration while still jumping over holes in the array.

It also remains future work to update the compiler implementation to support integers of arbitrary size. The current implementation supports signed 32-bit integers only, and overflows are not detected during label or condition evaluation. Similarly, it remains future work to define what happens when one attempts to divide by zero. Currently, division by zero is possible when computing label expression values in rule RHSs and also when evaluating rule conditions. It is possible that one could simply disallow the division operator in conditions, and implicitly have a condition inserted that ensures that division by zero never happens when evaluating RHSs, though its not obvious if this is a reasonable restriction, or if it is better to fail loudly. Another solution could be to introduce a new special runtime error state into the semantics of the language, and insist that any division by zero that occurs to runtime must transition straight to this state and terminate.

There remains an issue with edge conditions appearing negated in boolean expressions, and other complex expressions. This issue is documented on the GitHub issue tracker\footnote{\url{https://github.com/UoYCS-plasma/GP2/issues/10}}, and affects both the \texttt{legacy} and \texttt{master} versions of the GP\,2 Compiler. For example, the following will result in the compiler producing incorrect code: \texttt{not (edge(n0, n1) and edge(n1, n0))}. We are also aware of an issue with \texttt{try} statements within the guard of an \texttt{if}\footnote{\url{https://github.com/UoYCS-plasma/GP2/issues/28}}.

Finally, ongoing current research is exploring what classes of graph algorithms can be implemented in linear time in GP\,2 using the current compiler. Bak showed (undirected) DFS of graphs of bounded degree and bounded components can be performed in linear time \cite{Bak-Plump12a,Bak15a}, and also 2-colouring of such graphs \cite{Bak-Plump16a}. Due to this report, we know that we can sometimes perform reduction algorithms that don't limit the growth of the number of graph components in linear time, and Campbell, Courtehoute and Plump recently showed that GP\,2 can recognise trees and topologically sort DAGs of bounded degree in linear time \cite{Campbell-Courtehoute-Plump19b}.

\appendix
\chapter{Usage Documentation} \label{appendix:usage}

In this Appendix, we document the usage of the \texttt{legacy} and \texttt{master} branches of the GP\,2 Compiler, as they stand at 20th September 2019. We also take note of the unofficial Dockerized version of GP2, \enquote{GP2I}\footnote{\url{https://gitlab.com/YorkCS/Batman/GP2I}}, and the supporting ecosystem. At time of writing, the GP2 Editor\footnote{\url{https://github.com/UoYCS-plasma/GP2-editor}} has not been updated to use the \texttt{master} version of the compiler, and should be used with the \texttt{legacy} branch.


\section{Legacy Compiler}

The \texttt{legacy} branch specifies the state of the compiler before our modifications. We keep it available so users may see for themselves the differences in the prior and new compiler versions.

Installing the GP\,2 legacy compiler may be done with the following sequence of commands on any standard Linux or Mac computer. If any programs needed are not installed, install them.

\begin{lstlisting}
  git clone -b legacy https://github.com/UoYCS-plasma/GP2.git
  cd GP2/Compiler
  ./configure
  make
  sudo make install
\end{lstlisting}

A number of flags are available for the GP\,2 legacy compiler. In order to validate a given program, a user should write

\begin{lstlisting}
  gp2 -p <program_file>
\end{lstlisting}

where \lstinline{<program_file>} is the path of the program to be verified. To validate a rule on its own, the rule should be written in its own file and the command

\begin{lstlisting}
  gp2 -r <rule_file>
\end{lstlisting}

should be executed. To validate a graph, the command

\begin{lstlisting}
  gp2 -h <graph_file>
\end{lstlisting}

should be used. Finally, if one wishes to compile a given program, the user should enter the command

\begin{lstlisting}
  gp2 [-c] [-d] [-m] [-l <rootdir>] [-o <outdir>] [--max-nodes <MAX>] [--max-edges <MAX>] <program_file>
\end{lstlisting}

A number of options are available, as can be seen. These are as follows:

\begin{itemize}[itemsep=-0.7ex,topsep=-0.7ex]
  \item \lstinline{-c}: Use graph copying instead of undoing on the stack.
  \item \lstinline{-d}: Compile with GCC debugging flags, if you should wish to use a debugger on your program.
  \item \lstinline{-m}: Compile with root reflecting matches.
  \item \lstinline{-l <rootdir>}: Specify the root directory for the installed compiler.
  \item \lstinline{-o <outdir>}: Specify the path where the compiled files should be placed.
  \item \lstinline{--max-nodes <MAX>}: Specify the maximum number of nodes the program's graph parser can handle.
  \item \lstinline{--max-edges <MAX>}: Specify the maximum number of edges the program's graph parser can handle.
\end{itemize}

The latter two options are necessary due to issues noted in the graph parser of the legacy compiler.

The compiler will, when given the necessary program file in this manner, produce a number of C source and header files, along with a Makefile at the specified output directory. When this is done, the user may simply run \lstinline{make} in this directory to generate the final executable, called \lstinline{gp2run}. To run this program on a graph, simply enter the \lstinline{/tmp/gp2} directory and execute

\begin{lstlisting}
  ./gp2run <graphdir>
\end{lstlisting}

The output graph will be written in the same path. To delete all the generated files, simply run \lstinline{make clean}; ensure nothing else is in the current path when doing this.


\section{New Compiler}

The new compiler (on the \texttt{master} branch) may be installed by the following means:

\begin{lstlisting}
  git clone https://github.com/UoYCS-plasma/GP2.git
  cd GP2/Compiler
  ./configure
  make
  sudo make install
\end{lstlisting}

Usage is identical to the legacy version of the compiler. However, some flags have changed:

\begin{itemize}[itemsep=-0.7ex,topsep=-0.7ex]
  \item \lstinline{-c}, \lstinline{--max-nodes}, and \lstinline{--max-edges} have been removed.
  \item \lstinline{-f}: Compile a program in fast shutdown mode, to avoid freeing memory at termination.
  \item \lstinline{-g}: Compile a program with minimal garbage collection. This requires fast shutdown to be enabled.
  \item \lstinline{-n}: Compile a program without graph node lists, using arrays instead.
  \item \lstinline{-q}: Compile a program quickly, without optimisations.
  \item \lstinline{-l <libdir>}: Specify the directory for the \textbf{library} code.
\end{itemize}

Also, instead of running \lstinline{make} to compile the generated C code, one must now run \lstinline{./build.sh}.


\section{Dockerized Version} \label{sec:docker}

The purpose of the GP2I is to provide a simple Dockerized interface that can be run on almost any AMD64 architecture Linux or MacOS, without needing to install or configure any software, other than Docker. The interface is simple, and abstracts away from the fact there is actually a compiler. One simply specifies the input program and host graph, and GP2I will compile the program using GCC 9.2, and execute it on the host graph. The \texttt{legacy} tagged image corresponds to the \texttt{legacy} branch of the GP\,2 Compiler, and the \texttt{latest} tagged image corresponds to the \texttt{master} branch of the GP\,2 Compiler.

Running GP2I is simple. Simply choose the \texttt{tag}, and relative location of the GP\,2 program \texttt{prog} and input host graph \texttt{host}, as shown below.

\begin{lstlisting}
  docker run -v ${PWD}:/data \
    registry.gitlab.com/yorkcs/batman/gp2i:<tag> \
    <prog> <host>
\end{lstlisting}

It is also possible to specify compiler flags by setting the \texttt{GP2\_FLAGS} environment variable.

\begin{enumerate}[itemsep=-0.7ex,topsep=-0.7ex]
\item On success, the output graph is written to \texttt{stdout} and the process exits with code \texttt{0}.
\item On compilation error, such as due to an invalid program, the details are written to \texttt{stderr}, and the process exits with code \texttt{1}.
\item On program error, such as an invalid host graph or program evaluating to \texttt{fail}, the details are written to \texttt{stderr}, and the process exits with code \texttt{2}.
\end{enumerate}

There also exists \enquote{GP2I as a service}. That is, we have a gRPC\footnote{\url{https://grpc.io/}} interface and server implementation that allows client implementation to execute GP\,2 programs on host graphs, and receive the program execution time back, along with the output graph or an error message. Originally developed in Summer 2018, there have been minor modifications since, not least, to support the new GP\,2 Compiler. The current interface is provided in Figure \ref{fig:gp2i-proto}.

\begin{figure}[h!]
  \vspace{0.4em}
\begin{lstlisting}[language=protobuf3,style=protobuf]
syntax = "proto3";

message Request {
  string program = 1;
  string graph = 2;
}

message Response {
  oneof payload {
    string graph = 1; // the output graph
    string error = 2; // an error message
  }
  uint32 time = 3; // execution time
}

service GP2I {
  rpc Interpret (Request) returns (Response);
}
\end{lstlisting}
  \centering
  \vspace{-0.6em}
  \caption{The \texttt{gp2i.proto} Interface Description}
  \label{fig:gp2i-proto}
  \vspace{0.4em}
\end{figure}

\chapter{Software Testing} \label{appendix:tests}

In this Appendix, we give an account of the integration tests\footnote{\url{https://gitlab.com/YorkCS/Batman/Tests}} used to give some confidence in the correctness of both the \texttt{legacy} and \texttt{master} GP\,2 Compiler. As of 20th September 2019, there are 140 test cases that are checked, divided into two test suites, \texttt{quick} and \texttt{slow}. The quick suite contains 138 quick running tests, designed to test the correctness of the compiler on both small, simple test cases, and small, edge cases. The slow suite contains 2 tests that are executed on massive input graphs, designed to check the compiler does not fall over when given very large input graphs.

The tests execute using the GP2I docker images, as described in Section \ref{sec:docker}. Root reflecting mode (Section \ref{sec:rootrefl}) is used to test the output graphs are correct up to isomorphism by means of the generated program in Figure \ref{fig:iso-gp2}, where \texttt{\$GRAPH} is set to the expected graph.

\begin{figure}[h!]
  \vspace{0.4em}
\begin{lstlisting}[style=plain]
Main = remove; if {r1, r2, r3, r4} then fail

remove()
$GRAPH
=>
[ | ]
interface={}

r1(x: list)
[ (n1, x) | ]
=>
[ (n1, x) | ]
interface={n1}

r2(x: list)
[ (n1(R), x) | ]
=>
[ (n1(R), x) | ]
interface={n1}

r3(x: list)
[ (n1, x # any) | ]
=>
[ (n1, x # any) | ]
interface={n1}

r4(x: list)
[ (n1(R), x # any) | ]
=>
[ (n1(R), x # any) | ]
interface={n1}
\end{lstlisting}
  \centering
  \vspace{-0.6em}
  \caption{The \texttt{iso.gp2} Pseudo-Program}
  \label{fig:iso-gp2}
  \vspace{0.4em}
\end{figure}

We have also setup special debug GP2I images that start the generated program using Valgrind\footnote{\url{http://valgrind.org/}} to check for memory leaks and other memory errors, such as control flow that depends on uninitialised memory, on the \texttt{master} version of GP2. Moreover, we run the quick tests using various different combinations of the compiler flags, although we don't run all the combinations through Valgrind, since fast shutdown mode (Section \ref{sec:fastshutdown}) intentionally does not free all allocated heap memory.

\chapter{Benchmarking Details} \label{appendix:benchmarking}

In this Appendix, we give a quick overview of the benchmarking software used in this report. Note that all benchmarks were run on a MacBook Pro (Retina, 15-inch, Mid 2015) with 2.5 GHz Intel Core i7 and 16GB RAM. We support both the new and legacy compilers. For our benchmarking, we set both the max nodes and max edges parameter to \(8388608\), and when benchmarking the new compiler, we enabled \enquote{fast shutdown mode}.

The benchmarking software uses gRPC, and in particular, the GP2I server described in Section \ref{sec:docker}. There are two additional gRPC services. A graph generation service, and a benchmarking service. The graph generation service communicates with the GP2I service in order to generate graphs using GP\,2 programs, and the benchmarking service communicates with the graph generation service to generate input graphs, and then with the GP2I service in order to time the execution of programs on input graphs.

There is a benchmarking CLI client that calls the benchmarking service with input programs, and graph generation parameters. The service then streams back progress information for the client to display, and then, once benchmarking is complete, streams the results for the client to display.

\begin{figure}[h!]
  \vspace{0.4em}
\begin{lstlisting}[style=plain]
syntax = "proto3";

message Range {
  uint32 step = 1;
  uint32 min = 2;
  uint32 max = 3;
}

message Config {
  string mode = 1;
  Range range = 2;
}
\end{lstlisting}
  \centering
  \vspace{-0.6em}
  \caption{The \texttt{config.proto} Interface Description}
\end{figure}

\begin{figure}[h!]
\begin{lstlisting}[style=plain]
syntax = "proto3";

import "config.proto";

message Graph {
  string graph = 1;
  uint32 nodes = 2;
  uint32 edges = 3;
}

service Gen {
  rpc Generate (Config) returns (stream Graph);
}
\end{lstlisting}
  \centering
  \vspace{-0.6em}
  \caption{The \texttt{gen.proto} Interface Description}
\end{figure}

\begin{figure}[h!]
  \vspace{0.8em}
\begin{lstlisting}[style=plain]
syntax = "proto3";

import "config.proto";

message Program {
  string name = 1;
  string content = 2;
}

message Setup {
  repeated Program programs = 1;
  Config config = 2;
  uint32 runs = 3;
}

message Progress {
  uint32 phase = 1;
  uint32 total = 2;
  uint32 complete = 3;
}

message Benchmark {
  uint32 nodes = 1;
  uint32 edges = 2;
  float rate = 3;
  float time = 4;
}

message Benchmarks {
  string program = 1;
  repeated Benchmark results = 2;
}

message BenchStep {
  oneof payload {
    Progress progress = 1;
    Benchmarks benchmarks = 2;
    string error = 3;
  }
}

service Bench {
  rpc Execute (Setup) returns (stream BenchStep);
}
\end{lstlisting}
  \centering
  \vspace{-0.6em}
  \caption{The \texttt{bench.proto} Interface Description}
\end{figure}

\begingroup
\sloppy
\cleardoublepage
\phantomsection
\addcontentsline{toc}{chapter}{Bibliography}
\printbibliography
\endgroup

\end{document}